\documentclass[]{aastex}
\usepackage{emulateapj5,pstricks,psfig}

\newcommand{\beqa}{\begin{eqnarray}}
\newcommand{\eeqa}{\end{eqnarray}}
\def\d{\rm d}
\def\hkpc{\ h^{-1}{\rm kpc}}
\def\hMsun{\ h^{-1}M_{\odot}}
\def\hMpc{\ h^{-1}{\rm Mpc}}

\def\lt{{\lambda}'}
\def\rs{r_s}

\def\hkpc{\ h^{-1}{\rm kpc}}
\def\hMsun{\ h^{-1}M_{\odot}}
\def\hMpc{\ h^{-1}{\rm Mpc}}
\def\gsim {\lower .1ex\hbox{\rlap{\raise .6ex\hbox{\hskip .3ex
        {\ifmmode{\scriptscriptstyle >}\else
                {$\scriptscriptstyle >$}\fi}}}
        \kern -.4ex{\ifmmode{\scriptscriptstyle \sim}\else
                {$\scriptscriptstyle\sim$}\fi}}}
\def\lsim {\lower .1ex\hbox{\rlap{\raise .6ex\hbox{\hskip .3ex
        {\ifmmode{\scriptscriptstyle <}\else
                {$\scriptscriptstyle <$}\fi}}}
        \kern -.4ex{\ifmmode{\scriptscriptstyle \sim}\else
                {$\scriptscriptstyle\sim$}\fi}}}
\def\beq{\begin{equation}}
\def\eeq{\end{equation}}

\def\cvir{c_{\rm {\tiny{v}}}}
\def\Rvir{R_{\rm v}}
\def\rvir{R_{\rm v}}
\def\Mvir{M_{\rm v}}
\def\mvir{M_{\rm v}}
\def\Vvir{V_{\rm v}}
\def\Dvir{\Delta_{\rm v}}

\def\be{\begin{equation}}
\def\ee{\end{equation}}
\def\se#1{\S\ref{sec:#1}}
\def\fig#1{Fig.~\ref{fig:#1}}
\def\equ#1{Eq.~(\ref{eq:#1})}
\def\ifm#1{\relax\ifmmode#1\else$\mathsurround=0pt #1$\fi}

\def\hkpc{\,h\ifm{^{-1}}{\rm kpc}}

\def\omm{\Omega_{\rm m}}
\def\oml{\Omega_{\Lambda}}

\def\lcdm{$\Lambda$CDM}

\newenvironment{inlinefigure}{ 
\def\@captype{figure} 
\noindent\begin{minipage}{0.999\linewidth}\begin{center}} 
{\end{center}\end{minipage}\smallskip}

\begin{document}
\slugcomment{{\em Astrophysical Journal, submitted}}

\lefthead{A UNIVERSAL ANGULAR MOMENTUM PROFILE FOR GALACTIC HALOS}
\righthead{Bullock et al.}

\title{A universal angular momentum profile for galactic halos}

\author{J. S. Bullock$^1$, A. Dekel$^2$, T. S. Kolatt$^2$,   
   A. V. Kravtsov$^{1,}$\footnotemark[0], 
A. A. Klypin$^3$, 
   C. Porciani$^2$, \& J. R. Primack$^4$}

\affil {$^1$Department of Astronomy, Ohio State University,
Columbus, OH 43210 USA, james@astronomy.ohio-state.edu}

\affil {$^2$Racah Institute of Physics, The  Hebrew University,
Jerusalem 91904, Israel}

\affil {$^3$Astronomy Department, New Mexico State University, Box 30001, Dept.  4500, Las Cruces, NM 88003 USA }

\affil {$^4$Physics Department, University of California, Santa
Cruz, CA 95064 USA}


\begin{abstract}


We study the angular-momentum profiles of a statistical sample of halos drawn 
from a high-resolution $N$-body simulation of the $\Lambda$CDM cosmology.
We find that the cumulative mass distribution of specific angular momentum $j$ in a halo 
of mass $M_{\rm v}$ is well fit by a {\it universal~} function,  
$M(<j) = M_{\rm v} \mu j/(j_0+j)$.  This profile is defined by one shape 
parameter ($\mu$ {\em or} $j_0$) in addition to the global spin parameter 
$\lambda$.  It follows a power-law $M(<j) \propto j$ over most of the mass, 
and flattens at large $j$, with the flattening more pronounced for small 
values of $\mu$ (or large $j_0$ at a fixed $\lambda$).  Compared to a uniform 
sphere in solid-body rotation, most halos have a higher fraction of their mass 
in the low- and high-$j$ tails of the distribution.  High-$\lambda$ halos tend 
to have high $\mu$ values, corresponding to a narrower, more uniform $j$ distribution.
The spatial distribution of angular momentum in halos tends to be cylindrical 
and is well-aligned within each halo for $\sim 80\%$ of the halos.  The more 
misaligned halos tend to have low-$\mu$ values.  When averaged over  
spherical shells encompassing mass $M$, the halo $j$ profiles are fit by  
$j(M) \propto M^s$ with $s=1.3 \pm 0.3$.  
We investigate two ideas for the origin of this profile.
The first is based on a revised version of linear tidal-torque theory combined 
with extended Press-Schechter mass accretion, and the second focuses on $j$ 
transport in minor mergers. 

Finally, we briefly explore implications of the $M(<j)$ profile
on the formation of galactic 
disks assuming that $j$ is conserved during an adiabatic baryonic infall. 
The implied gas density profile deviates from an exponential disk, with a 
higher density at small radii and a tail extending to large radii. 
The steep central density profiles may imply disk scale lengths that are 
smaller than observed.  This is reminiscent of the ``angular-momentum problem"
seen in hydrodynamic simulations, even though we have assumed perfect $j$ 
conservation.  A possible 
solution is to associate the central excesses
with bulge components and the outer regions with extended gaseous disks. 

\end{abstract}

\subjectheadings{cosmology --- dark matter --- galaxies: 
formation --- galaxies: structure}

\section{Introduction}
\label{sec:intro}

\label{sec:intro}
\footnotetext[0]{Hubble Fellow}

The origin of the distribution of mass and angular momentum in disk
galaxies is still an open issue, despite its long history.

The archetypical model of Eggen, Lynden-Bell \& Sandage (1962) 
provides a useful framework, in which a disk galaxy (the Milky Way) forms
by centrifugal support of a collapsing gas cloud.
Mestel (1963) added the assumption that the specific angular momentum
of each mass element, $j$, is conserved during the collapse,
and demonstrated that the final mass profile of the disk could
then be related to the initial mass and angular momentum distribution
of the gas.
The key is that the 
cumulative
mass with specific angular momentum less than $j$,
$M(<j)$, is preserved during the collapse. 
Crampin \& Hoyle (1964) (also Innanen 1966; Freedman 1970) then realized
that the observed exponential density profiles of disk galaxies in circular
motions are consistent with the angular momentum distribution of a 
hypothetical uniform sphere in solid body rotation,
\begin{equation}
M(<j) = M_{\rm tot}\left[1 - (1 - j/j_{\rm max})^{3/2}\right]
\label{eqt:iso}
\end{equation}
where $j_{\rm max} = \omega R^2$ with $\omega$ the fixed angular velocity
and $R$ the radius of the sphere, and $M_{\rm tot}$ the total mass
of the sphere.

Fall \& Efstathiou (1980) re-examined this question in the context of the 
more modern view according to which disk galaxies form 
from contracting 
gas within extended dark-matter halos (White \& Rees 1978). 
They added the assumption that the gas and dark
matter were initially well-mixed such that the distribution
of $j$ in the disk is equal to that of the halo.
Rather than assuming an angular momentum profile, they implicitly
deduced it from the constraint that
the final surface
density profile of the disk is exponential.
The implied disk scale radius, $r_{\rm d}$, 
is related to the total pre-collapse angular momentum
of the halo,  roughly $r_d \propto \lambda$.  Here, $\lambda$
is the dimensionless spin parameter (e.g. Peebles 1969)
\begin{equation}
        \lambda \equiv \frac{J\left|E\right|^{1/2}}{G M^{5/2}},
\label{eq:lambda}
\end{equation}
where $J$,$E$ and $M$ are the total angular momentum, energy and mass
of the system, and $G$ is Newton's constant.  

Angular momentum is presumably acquired by the dark matter (and gas)
 through tidal interactions with neighboring objects (Peebles 1969).
Barnes \& Efstathiou (1987) (following 
Efstathiou \& Jones 1979; Efstathiou \& Barnes 1983; 
Zel'dovich \& Novikov 1983) utilized cosmological N-body simulations
with a CDM power spectrum to make detailed predictions for
halo angular momentum structure and statistics.  
In particular they found
that the distribution of 
spin parameters is roughly log normal, with a median value of 
$\lambda \sim 0.05$.  They also confirmed that the angular momentum of
pre-collapse halos grows roughly linearly with time, as predicted by
linear tidal-torque theory (Doroshkevich 1970; White 1984).
Several numerical and analtyical 
investigations have followed, which confirmed the
nature of the $\lambda$ distribution, 
addressed the alignment of the spin
vectors between neighboring galaxies, the (lack of) dependence of halo
spin on environment, and the effect of major mergers on halo spin  (e.g.,
Frenk et al. 1988; Heavens \& Peacock 1988;
Zurek, Quinn \& Salmon 1988; Warren et  al. 1992;  
Catelan \&  Theuns 1996; Cole \& Lacey   1996; 
 Lemson \& Kauffmann 1997; Gardner 2000).

Armed with this global information about the initial state
of the dark matter and gas mixture, several authors have
used the above disk formation scenario to predict observable quantities.
Blumenthal et al. (1986) addressed the rotation curves of disk galaxies
in the context of the CDM hierarchical formation scenario,
assuming $\sim 10\%$ of the mass in the gas component 
(Blumenthal et al. 1984).   They pointed out that the non-dissipative dark halo
should react to the dissipative gas infall by considerable contraction,
and showed that this process can be approximated
using
an adiabatic
invariant.  This coupling between the halo and the disk
``conspires" to produce
a continuity between the disk-dominated and halo-dominated
regions of the rotation curves, as observed (Burstein \& Rubin 1985).

Flores et al. (1993) explored the dependence of the
final rotation curves on a range of assumed parameters, including
the measured spread in $\lambda$.  Dalcanton, Spergel, \& Summers (1997)
included a more realistic initial halo density profile
(Hernquist 1990) and explored the properties of both high
surface brightness and low surface brightness galaxies.  
Mo, Mao, \& White (1998a; 1998b, 1999) performed a similar examination 
within the context of various cosmologies, assuming that
the dark halos followed the density profile advocated by Navarro, Frenk, 
\& White (1996, 1997; NFW)
and compared their predicted galaxy properties 
to both local and high-redshift data.  van den Bosch (1998; 2000)
and  van den Bosch \& Dalcanton (2000)
extended the approach to include a bulge component and investigated 
the origin of the Hubble sequence of galaxy types and other aspects
of galaxy formation.  

The main commonality in all of these investigations is that the variation
in observable galaxy properties depends almost exclusively on the
initial total spin parameter of the halo hosting the galaxy. 
However, all of the predictions rest on the implicit 
assumption that, if specific angular momentum is conserved,
the initial angular momentum profiles of the halos
produce disks that are exponential in form.~\footnote{An
exception is Dalcanton et al. (1997), who
assumed that the initial $M(<j)$ profile was that of a uniform
solid body rotator, which 
produces disk
density profiles that are 
similar to exponential (Ryden \& Gunn 1987) but only over about
two disk scale lengths (Olivier, Primack, \& Blumenthal, 1991).}

Within this simple framework for disk formation,
the key missing ingredient is the actual
distribution of angular momentum in halos.
Partially motivated by this, we set out to measure the $j$ profiles
directly using a statistical sample of halos in a high resolution simulation,
and to estimate the intrinsic scatter in these distributions.
The long-term goal is to learn what the detailed form
of halo angular momentum profiles 
can teach us about the origin of galaxy types, 
disk surface brightness profiles, and bulge properties.

An additional strong motivation for this study comes from
the results of hydrodynamic simulations of disk galaxy formation in a 
cosmological context. Unlike the simplified, monolithic-collapse framework 
discussed above, the hydrodynamic treatments model gas cooling
within the hierarchical growth of dark halos, and typically
find that the 
resulting
disks are significantly smaller 
than real 
galactic disks
(Navarro \& Steinmetz 2000;
Weil, Eke, \& Efstathiou 1998; Navarro \&
Steinmetz 1997; Navarro, Frenk, \& White 1995). 
The problem apparently arises because
most of the mass is accreted through mergers with objects whose
gas component has already cooled and clumped, causing
a large fraction  of their
angular momentum to be 
transferred to the dark halo or
transported outwards.
Since angular-momentum transport is likely to always be from the inside out, 
the simple case in which $j$ is conserved during the collapse, which we can
study in detail with high accuracy, can serve as a useful limiting
case. This will provide a better basis for addressing the more
complex physical processes required for solving the angular-momentum
problem seen in the hydro simulations.

In \se{method} we describe 
our methods,
including the simulation and halo 
finding, the measurement of the spin parameter, the associated errors,
and the measurement of the angular-momentum profile.
In \se{mass-dist} we discuss the measured 
(universal) mass distribution
of angular momentum 
in halos.
In \se{spatial-dist} we address the spatial distribution 
of angular momentum,
its alignment, cylindrical symmetry, and profile in spherical shells.
In \se{origin} we use both tidal-torque theory and
an estimate of angular momentum  transfer due to minor mergers to
explore the origin of the angular-momentum profile.
In \se{disk} we briefly describe the implications for the disk 
surface density profile and the ``angular momentum catastrophe".
We discuss our results and conclude in \se{conc}.  
An Appendix is devoted to 
testing our results with a simulation
having $8$ times the mass resolution, and investigating
$M(<j)$ profiles of halos at high redshift.

\section{Method}
\label{sec:method}

\subsection{Halos in the ART simulation}

Large cosmological N-body simulations have reached the
stage where detailed structural properties of many dark-matter halos
can be resolved simultaneously.  
A method which allows the required force and mass resolution 
is the Adaptive Refinement Tree (ART) method (Kravtsov, Klypin,
\& Khokhlov 1997)
which implements successive refinements of the spatial grid and
time steps in high density environments.  The simulations based on 
the ART code provide a statistical sample of halos, 
covering a wide mass range, with the resolution appropriate for 
studying angular-momentum profiles.

We have used the ART code to simulate the evolution of collisionless
dark matter within the ``standard"
low-density flat $\Lambda$CDM model ($\Omega_{\rm m}=0.3$, 
$\Omega_{\Lambda}=0.7$, $h=0.7$, and $\sigma_8=1.0$ at $z=0$).  
The  simulation followed the trajectories of $256^3$  particles within a 
periodic box of comoving size $60\hMpc$ from redshift $z = 40$ to the
present.   A basic $512^3$ uniform grid was used, and six
refinement levels were introduced in the regions of highest density, implying
a dynamic range of $\sim 32,000$.  The formal resolution of the simulation
is thus $f_{\rm res} =1.8 \hkpc$, and the mass per
particle is $m_{p} = 1.1 \times 10^{9} \hMsun$.

Our halo finding algorithm (Bullock et al. 2000) is based on the Bound 
Density Maxima technique (Klypin \& Holtzman 1997) 
and has been specifically designed to 
identify halos (and subhalos) in such a high resolution simulation.
In the current study of angular momentum, we limit ourselves to
`distinct' 
halos, which do not reside within a larger host halo,
in the mass range $10^{12}-10^{14} \hMsun$.  In the Appendix, we
use a higher resolution simulation to extend our mass range down
to $\sim 10^{11} \hMsun$.

A brief description of the halo finder is as follows:
After finding all the maxima in the smoothed density field of the simulation, 
we unify overlapping maxima, define a minimum number of particles per 
halo (say 50), and iteratively find the center of mass of a sphere about
each of the remaining maxima.
We compute the spherical density profile about each center
and identify the halo virial radius $\rvir$
inside which the mean overdensity has dropped to a value $\Dvir$,
based on the top-hat spherical infall model.
For the family of flat cosmologies ($\omm+\oml=1$),
the value of $\Dvir$ can be approximated by (Bryan \& Norman 1998)
$\Dvir \simeq (18\pi^2 + 82x - 39x^2)/(1+x)$,
where $x\equiv \Omega_m(z)-1$.
In the \lcdm\ model used in the current paper, $\Dvir$ varies from
about 180 at $z\gg 1$ to $\Dvir\simeq 340$ at $z=0$.

We then fit each halo density profile with a universal functional form.
We adopt the NFW profile,
\begin{equation}
\rho_{\rm{NFW}}(r) = \frac{\rho_s}{(r/\rs)\left(1+r/\rs\right)^2},
\label{eq:nfw}
\end{equation}
with the two free parameters $\rs$ and $\rho_s$.
An equivalent pair of parameters is, for example, the virial mass $\mvir$
and the concentration parameter $\cvir \equiv \rvir/\rs$.
Using this fit, we iteratively remove unbound particles from each modeled
halo, and unify every two halos that overlap in their $\rs$ and are
gravitationally bound.\footnote{Removal of unbound particles is a new
feature allowed by the high resolution
of our simulation; it could not have been properly
implemented in earlier studies.}
The modeling of the halos with a given functional form allows us
to assign to them characteristics such as a virial mass and radius,
and to estimate sensible errors for these quantities.

The halo finding is 100\% complete for halos of more than $\sim 150$ particles,
$M \geq 1.5 \times 10^{11} \hMsun$ (see Sigad et al. 2000).
For our purpose here, we limit the sample to halos of more than $\sim 1000$
particles.

\begin{inlinefigure} 
\resizebox{\textwidth}{!}{\includegraphics{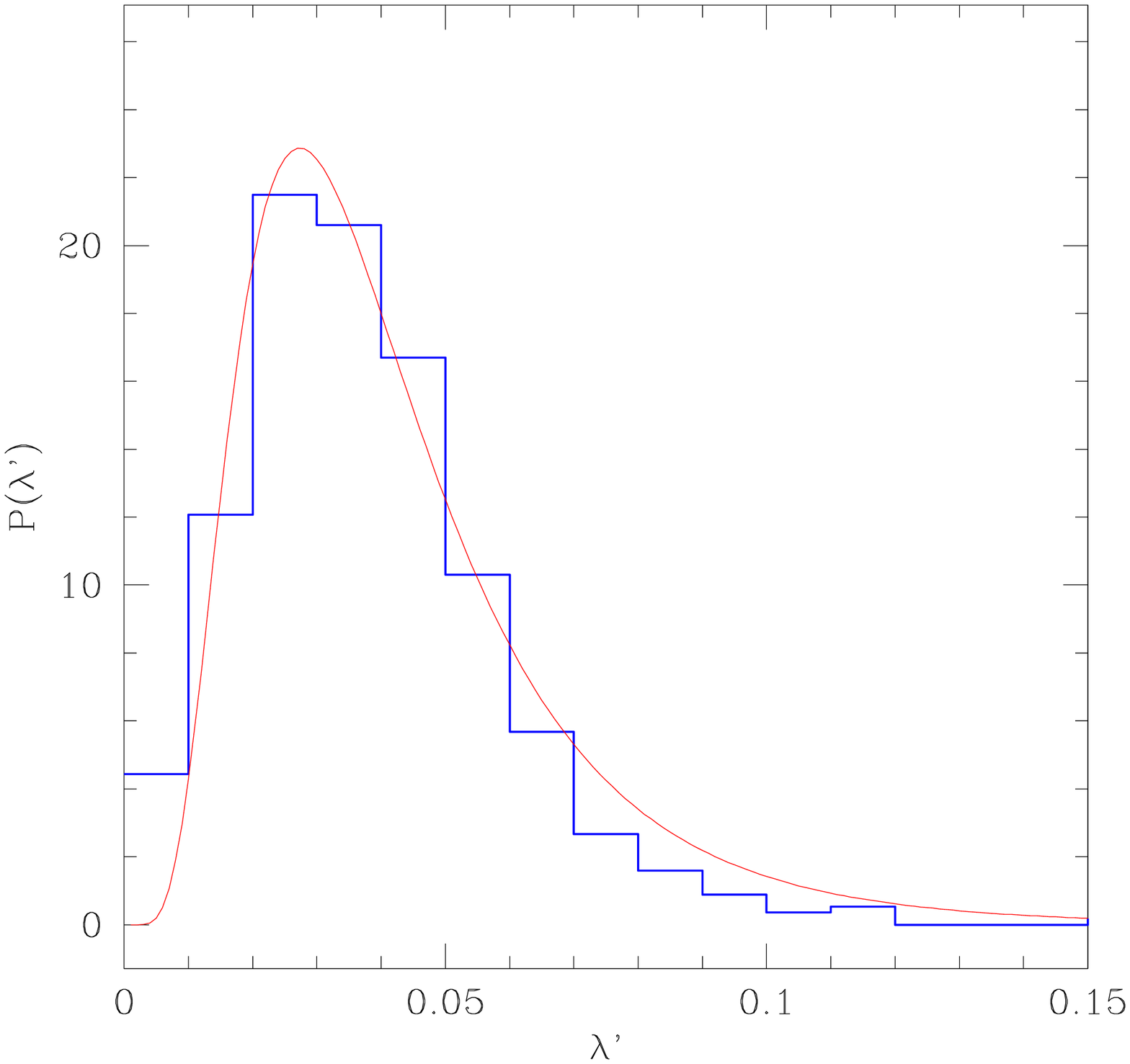}} 
\figcaption{The distribution of halo spin parameter $\lt$ (histogram),
 compared to a log-normal distribution about $\lt_0 = 0.035 \pm 0.005$
 with a width $\sigma = 0.50 \pm 0.03$.  
\label{fig:lt_dist}} 
\end{inlinefigure} 

\subsection{Angular-momentum errors} 

A first step towards measuring the distribution of specific angular momentum
within each of the halos is measuring the {\it global}
spin parameter of each halo.
The angular momentum of a halo of $N$ particles is defined by
\begin{equation}
        \textbf{J} = m_i \sum_{i=1}^{N} \bf{r}_{i} \times \bf{v}_{i},
\end{equation}
where $\bf{r}_{i}$ and $\bf{v}_{i}$ are the position 
and velocity 
of the $i$th particle with respect to the halo center of mass.  
In principle, given the spin parameter $\lambda$, the value of the 
global 
specific angular momentum, $J/M$, can be determined by using 
 an assumed energy content for the halo in \equ{lambda}.
In practice, however, this is not a straightforward procedure.
For example, the energy of a halo in a crowded region is somewhat ambiguous
because it depends on the environment.
A related difficulty arises when only a sub-volume of the
virial sphere is concerned, e.g., within the cooling radius.
We therefore define an alternative and more practical spin parameter by
\begin{equation}
	\lt \equiv \frac{J}{\sqrt{2}MVR}, 
\label{eq:lt}
\end{equation}
given the angular momentum $J$ inside a sphere of radius $R$ 
containing mass $M$,
and where $V$ is the halo circular velocity at radius $R$, $V^2=GM/R$.
This spin parameter reduces to the standard $\lambda$ when
measured at the virial radius of a truncated
singular isothermal halo.
The value of $\lt$ turns out to be robust to the choice of an
outer radius; we use below the virial radius $\rvir$,
but we find a very similar distribution of $\lt$ values when using $\rvir/2$
instead.  This allows modelers to freely scale our results using any 
desired outer radius appropriate for the problem at hand.

\begin{inlinefigure} 
\resizebox{\textwidth}{!}{\includegraphics{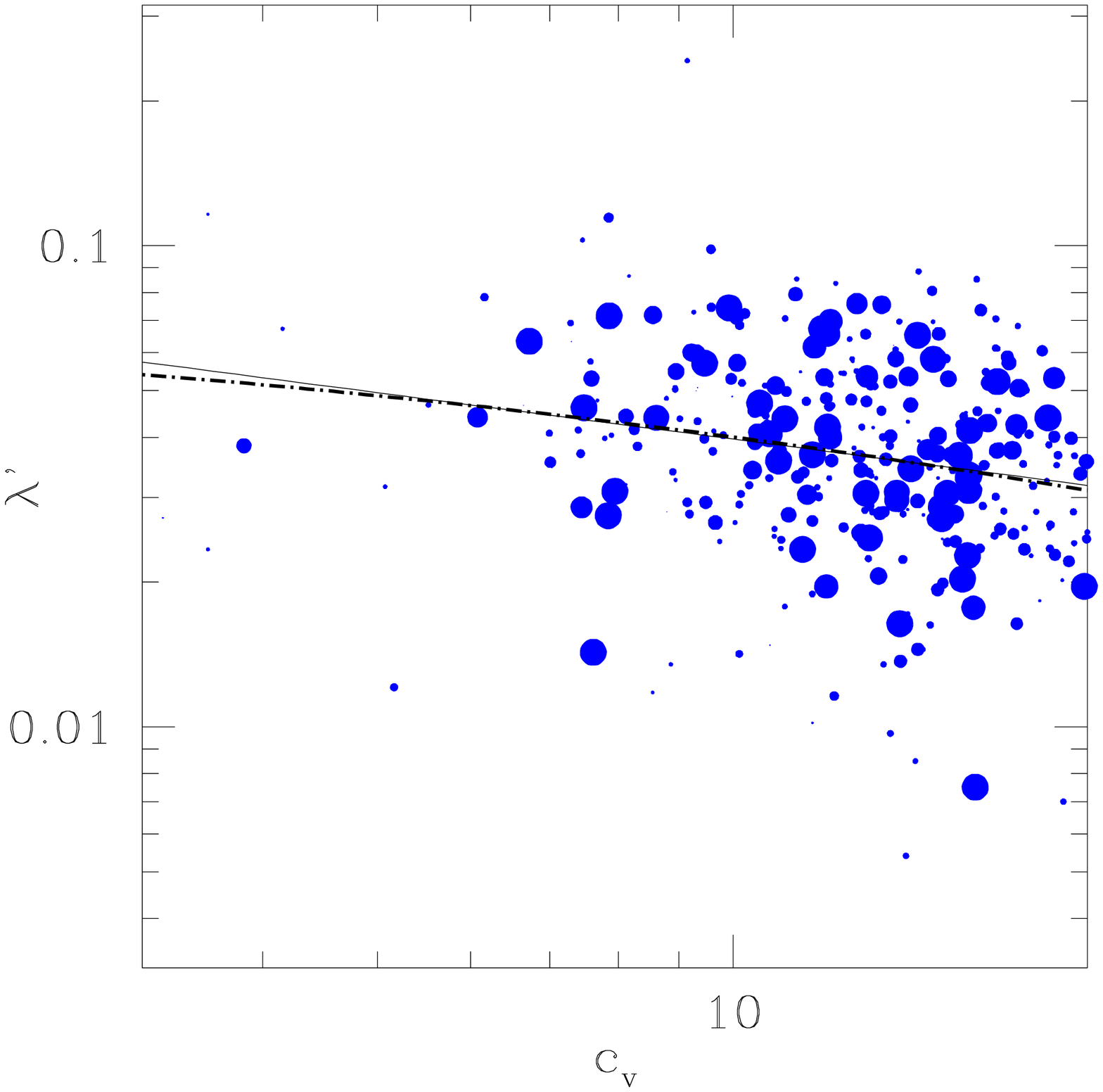}} 
\figcaption{
Halo spin parameter $\lt$ versus concentration $\cvir$.
  The symbol size is inversely proportional to the error on
  $\cvir$.  Shown are the linear regression line
  of $\lt$ on $\cvir$ (solid), and the relation expected
  assuming that the standard spin parameter $\lambda$ is
  independent of $\cvir$ (dot-dashed).
\label{fig:lt_vs_c}} 
\end{inlinefigure} 

The spin parameters $\lt$ and $\lambda$ are in fact very similar 
for typical NFW halos.  They are related by 
$\lt = \lambda |E_{SIS}/E_{NFW}|^{1/2}  \simeq \lambda  f(\cvir)^{-1/2}$,
where $E_{SIS}$ and $E_{NFW}$ are the energies of virialized halos
with isothermal and NFW density profiles respectively, and
$f(\cvir) \simeq [2/3  + (\cvir/21.5)^{0.7}]$ (Mo et al. 1998)
\footnote{The exact expression is $f(\cvir) = 
0.5 \cvir [ (1+\cvir)^2 - 1 - 2(1+\cvir)\ln(1+\cvir)]/[\cvir - 
(1+\cvir)\ln(1+\cvir)]^2$.}.
The function $f(\cvir)$ is about unity for typical concentrations,
$\cvir \sim 10$.

The distribution of $\lt$ over the $\sim 500$ halos in our sample
is shown in \fig{lt_dist}. It is well fit by a log-normal
distribution, 
\begin{equation}
        P(\lt) = \frac{1}{\lt\sqrt{2\pi}\sigma}\exp\left(-\frac
        {\ln^2(\lt/\lt_{0})}{2\sigma^2}\right),
\end{equation}
with best fit values $\lt_{0} = 0.035 \pm 0.005$ and
$\sigma = 0.5 \pm 0.3$.
Not surprisingly, the distribution of $\lt$ is very similar to the known
distribution of $\lambda$ (Barnes \& Efstathiou 1987).  The
distribution of $\lambda$ values for our halos has best
fit values $\lambda_{0} = 0.042 \pm 0.006$ and
$\sigma = 0.5 \pm 0.35$.

To highlight a small difference between $\lt$ and $\lambda$,
\fig{lt_vs_c} displays the values of $\lt$ versus $\cvir$ for our
sample of simulated halos, and shows a weak correlation. 
Given that $\lambda$ is known to be uncorrelated with $\cvir$ (NFW), 
this correlation arises from $f(\cvir)$ in the
above relation between the two spin parameters (as demonstrated in the
figure)~\footnote{In agreement with the NFW result, we also
find no correlation between $\cvir$ and the traditional spin
parameter $\lambda$.}.

\begin{inlinefigure}     
\resizebox{\textwidth}{!}{\includegraphics{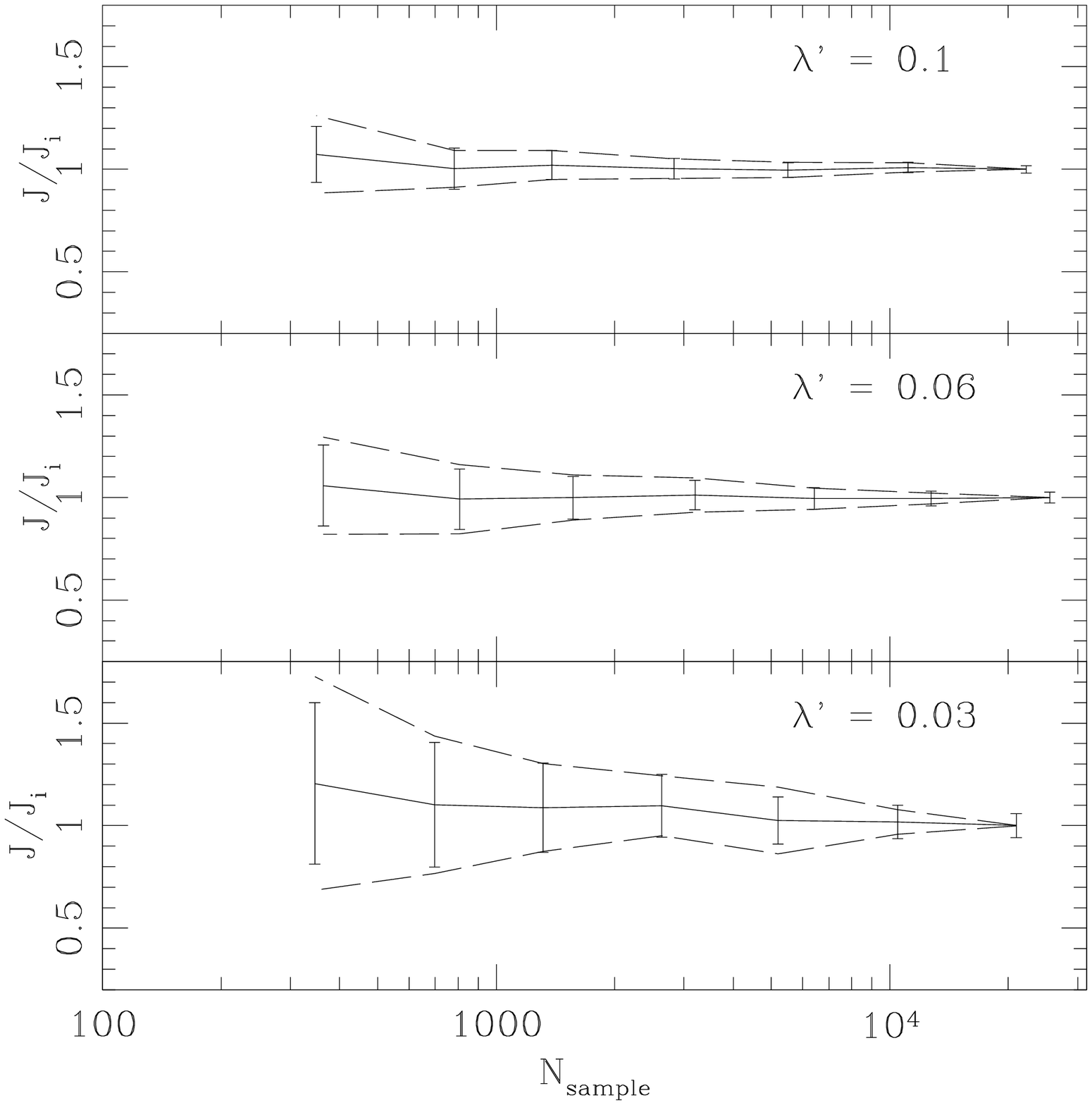}}  
\figcaption{
Estimate for the error in $J$.
The average (solid line) and standard deviation (dashed line) of $J/J_i$ 
as a function of the number of particles in the random subsamples.
    \label{fig:er}}
\end{inlinefigure}

The error on the measurement of $\bf{J}$ in the simulations
is of primary importance for the interpretation of the results
and, here, for the specific task of fitting a functional form 
to the angular-momentum profile.
Obtaining a precise measurement of $\bf{J}$ in halos is difficult
because the signal-to-noise is relatively small due to the fact
that the local 
coherent velocity of the halo contributing
to the angular momentum is small compared the velocity dispersion.
The variable $\lt$ itself roughly characterizes the signal to noise 
(per particle) at any radius $R$, \equ{lt}, 
which is typically at the level of a few percent.

In order to obtain a crude estimate of the error, which is probably an
upper limit, we singled out several large halos with $N_i$ simulated
particles ($> 2\times 10^4$), 
and measured their angular momentum using a series of sparsely sampled 
realizations of $N$ particles down to $1\%$ of $N_i$, with
100 random realizations for each value of $N$.
The average and standard deviation of the measured value $J$
relative to the original value obtained with all the particles, $J_i$,
as a function of $N$, are shown in \fig{er} for three halos with
different values of $\lt$.
We find that for a subsample of $N$ particles
the standard deviation $\sigma_J$ scales like
\begin{equation}
\frac{\sigma_J}{J} = \sqrt{1/N + 1/(25 \lt^{2}N)} 
\simeq \frac{0.2}{\lt \sqrt{N}}.
\label{eq:er}
\end{equation}
This scaling relation is shown as error bars in \fig{er},
and it fits rather well the measured scatter.
The first term in \equ{er} corresponds to the general Poisson 
sampling 
error and the second term reflects the noise due to velocity dispersion.
The second term typically dominates because $\lt$ is small,
and thus the spin parameter itself directly influences our
ability to measure $J$ accurately, with the number of particles needed
for a fixed accuracy scaling like $\lt^{-2}$.  For 
$\lambda' = 0.03$, even $\sim 50\%$ accuracy requires $N \sim 150$.

The above estimate of the error probably 
tends to overestimate the true error 
because it includes small-scale noise that is not present when a smaller 
number of particles is used in the 
actual 
dynamical simulation of the halo.
On the other hand, our procedure above does not mimic the 
error made in following the true dynamics of the smooth halo when a smaller 
number of particles is used ---
an effect which tends to cause an underestimate of the error.
An accurate evaluation of the error in $J$ requires a direct comparison
of the same halo as simulated with different resolutions, and is clearly 
sensitive to 
the $N$-body code used. In our analysis here we
limit ourself to the crude estimate, \equ{er}.  In the Appendix
we use a simulation with $8$ times the mass
resolution in order to check for systematic differences in
the derived $M(<j)$ profiles, and find none.
This analysis gives us 
confidence that mass resolution
does not affect our results in any severe way.

\subsection{Measuring the angular-momentum profile} 
\label{sec:measure_prof}
Our goal is to determine how the angular momentum is distributed
in the halo. We wish to compute the mass distribution of specific
angular momentum, $M(<j)$, and also the spatial distribution of $j$.  

We first compute the total $\bf{J}$ for each halo, and let it
define the $z$ direction.
Then we subdivide the spherical volume of each halo
into many spatial cells, each containing many particles,
and compute the specific angular momentum in each cell,
projected along the $z$ direction.
In general we use the symbol $j$ 
to refer to this $z$-projected angular momentum.  In some
instances we also investigate the non-projected, absolute magnitude
of angular momentum in each cell, which we refer to as $|j|$.
When comparisons between the two measurements are made, we 
explicitly add the subscript $z$ to the projected value, $j_z$. 

The cell geometry and spatial distribution 
were
designed
to maximize the number of particles in each cell while at the same time
sampling as much of the $j$ distribution as possible.
The vector character of angular momentum naturally 
introduces a preferred
axis, convolved with the general spherical symmetry of the mass distribution.
The cells are defined using the usual spherical coordinates
($r$,$\theta$,$\phi$) about the halo center.
Each of our cells spans the full $2\pi$ range in $\phi$,
and they span the range of ($r/R_{\rm v}$,\, $\sin \theta$) 
from ($0,0$) to ($1,1$).  
The radial shells are spaced such that each contains the
same number of particles.
The shells are 
then divided into three azimuthal cells of equal volume between
$\sin \theta =0$ and $1$. 
Positions with the same $r$
$\sin \theta$ above and below the equatorial plane belong to the same cell.
For all the $M(<j)$ profiles studied in detail and presented in the
figures below, we adjusted the number of radial bins such that 
each cell contains roughly $N_{p}$ particles, where 
$N_p = 500$ if $\Mvir > 5.5 \times 10^{12} \hMsun$ and $N_p = 0.1 \Mvir/ m_p$
for $\Mvir < 5.5 \times 10^{12} \hMsun$.

We construct $M(<j)$ profiles for each halo by ranking
the cells by their $j$ values, and counting 
the cumulative mass in cells with angular momentum less than $j$.
Because $j$ corresponds to a projected component, 
it is possible for a cell to have a negative value of $j$. 
The anti-alignment
needed for a cell to have a negative $j$ value is rare, but when
one occurs, we remove the cell
completely from the ranked list used to construct the $M(<j)$ profile.
About $5 \%$ of halos have a significant amount of their total
mass ($ \ge 10\%$)  contained in negative $j$ cells.  We do
not consider these  anti-aligned halos in our $M(<j)$ 
analysis.\footnote{We include all halos when we present the
distribution of half-mass alignment
cosines in \S4.1}
 
We assign an error to the $j$ in each cell using \equ{er}.
Every halo analyzed has $N>1000$ particles within its
virial radius, corresponding to $M_{\rm v} \ga 10^{12} \hMsun$.
We have $\sim 600$ halos that meet our requirements.  In practice,
we determined the shape of the functional form which fits $M(<j)$
by examining the group of ($\sim 200$) halos with 
$N>6000$ particles, for which the profiles are determined with
greater accuracy.
We extended the mass range down to $1000$ particles in order
to determine whether more typical galactic-sized halos 
have a similar profile.  Although these smaller halos have 
larger errors on their fit parameters, they seem to obey
the same $M(<j)$ profile.

The main limitation of this analysis is mass resolution.
Because we need $\ga 100$ particles per cell in order
to obtain a reasonable measure of $j$,
our determination of $M(<j)$ is only reliable down to
 $\sim 10\%$ for $\Mvir < 5.5 \times 10^{12} \hMsun$ and 
$\sim (5 \times 10^4 m_p/\Mvir) \%$ for $\Mvir > 5.5 \times 10^{12} \hMsun$.

\begin{figure*}[ttt]
\centerline{\includegraphics[width=0.95\linewidth]{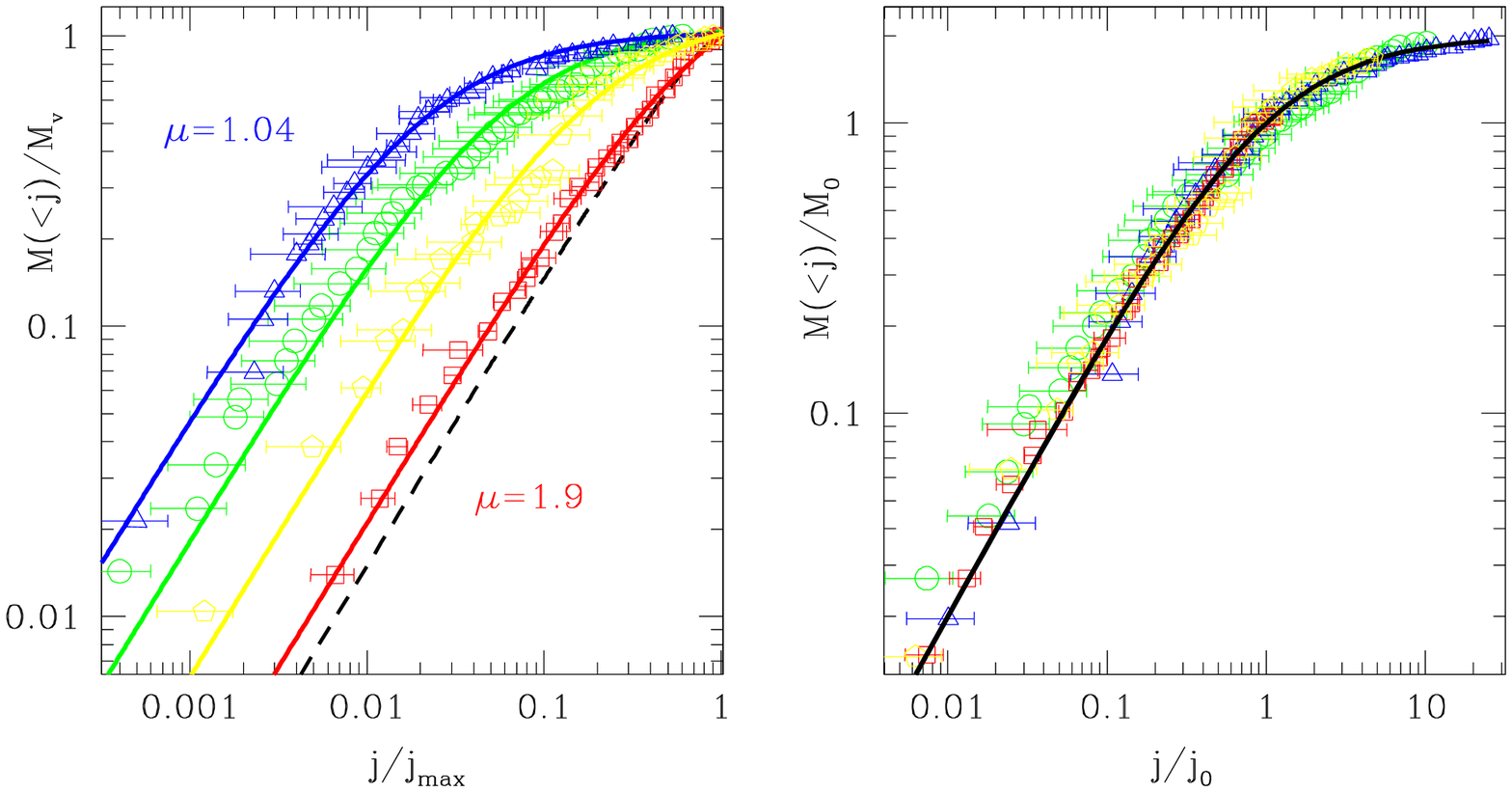} }
  \caption{ 
Mass distribution of specific angular momentum
in four halos spanning a range of $\mu$ values from 1.04 to 1.9.
Symbols and errors correspond to the ranked $j$ measurements in cells, 
while the curves are the functional fits, 
$M(<j) = \Mvir \mu j/(j_0 + j)$.
(a) All profiles 
are normalized to coincide at $\Mvir$, where $j=j_{\rm max}$. 
The value of $\mu$ measures the relative extent of the power-law regime
until it bends over.
Shown for comparison is the distribution for
a uniform sphere in solid-body rotation (dashed line).
(b) All profiles are normalized to coincide at $j_0$ and on top of
the universal profile (curve). 
The value of $\mu$ now correlates with the uppermost point, $j_{\rm max}/j_0$,
along the universal curve.
}\label{fig:examples} 
\end{figure*} 

\section{Mass distribution of angular momentum}
\label{sec:mass-dist}

\subsection{A universal angular-momentum profile}

We find that the specific angular-momentum profiles 
of halos in the simulation are well-fit by the
following 
two parameter function: 
\begin{equation}
M(<j) = M_{\rm v} \frac{\mu j}{j_0 + j}, \quad \mu > 1.
\label{eq:fit}
\end{equation}
The profile has an implicit maximum specific angular
momentum $j_{\rm max} = j_0/(\mu - 1)$.
It is  roughly a power-law for $j \la j_0$,
and flattens out for $j \ga j_0$. As can be seen
by its relation to $j_{\rm max}$, 
the quantity $\mu$ ($>1$) acts as a shape 
parameter: for 
$\mu \gg 1$, $M(<j)$ is a pure power law, while 
$\mu \rightarrow 1$ means that only half the mass falls within the
power-law regime and the bend is pronounced. 

The pair of parameters $\mu$ and $j_0$ fully defines 
the angular momentum 
distribution of the halo. 
The global spin parameter is related to $\mu$ and $j_0$ via
\begin{equation} 
j_0\, b(\mu) = \sqrt{2} V_{\rm v} R_{\rm v}\, \lambda' \,, 
\label{eq:par}
\end{equation}
where
\begin{equation} 
 b(\mu) \equiv \int_0^1 \frac{m}{\mu - m} dm =
-\mu\, \ln(1-\mu^{-1}) -1 . 
\label{eq:bmu} 
\end{equation}
Once the value of $\lt$ is known, Eq.~\ref{eq:fit} is a one-parameter
fitting function.  In fitting the $M(<j)$ profiles for each halo,
we first measured $\lt$ and 
then 
obtained the best fit $M(<j)$ with the
constraint that $\mu$ and $j_0$ return the measured $\lt$.
The pair $(\lt,\mu)$ is perhaps the most useful pair of parameters
for characterizing the halo angular momentum
because $\lt$ is the conventional global measure and $j_0$
can be explicitly determined from $\lt$ and $\mu$ via
\equ{par}, while $\mu$ cannot be determined explicitly from the
other two parameters without iteratively solving 
\equ{bmu}. 
Alternatively, we provide the following approximation for the 
iterative solution of \equ{bmu}: 
\beq
\mu(b) \simeq 1 + \frac{0.5}{[\exp(1.09 b) -1 ]},
\eeq
which is accurate to 
within 
$8\%$ in $\mu - 1$ over 
the range $\mu -1 = 0.01 - 10$.

Figure~\ref{fig:examples} shows $M(<j)$ examples for several
of our halos.  In the left panel we have normalized the profiles
by $\Mvir$ and $j_{\rm max}$ in order to illustrate how
different values of $\mu$ affect the distribution.
As $\mu$ approaches
its minimum value of $1$, a larger fraction of the halo mass
is spinning slowly compared to $j_{\rm max}$. Larger values of
$\mu$ imply a more 
uniform 
$j$ distribution.
The right panel shows the same four halos, now
normalized to their best-fit $j_0$ and $M_0$ values,   
where $M_0 = M(<j_0) = \mu \Mvir / 2$.  Notice
how remarkably the halos follow the characteristic functional shape.

The spread in profile 
shapes can be described by the distribution of 
$\mu - 1$ ($= j_0/j_{\rm max}$).
Figure~\ref{fig:muhist} shows the distribution of 
$\log_{10}(\mu -1)$ 
for all of our halos (histogram) along with a Gaussian
distribution with the same mean and standard deviation:
$\langle \log_{10}(\mu-1) \rangle =-0.6$, $\sigma = 0.4$.  The implied $90 \%$ range
is $\mu -1 \simeq 0.06 - 1.0$.

\subsection{Correlations between parameters}

Although the spin parameter $\lt$ and the shape parameter $\mu$  
clearly measure two different aspects of the angular-momentum distribution, 
they are not necessarily uncorrelated. Figure~\ref{fig:mu.vs.lambda} 
shows the joint distribution of these two parameters for our simulated halos.
We observe a weak but significant linear correlation between 
$\log (\mu - 1)$ and $\log \lt$, with 
a Pearson correlation coefficient of $r \simeq 0.23$ (corresponding to
a probability $p \sim 10^{-6}$ for no correlation).
Thus, high-spin systems tend to have more evenly distributed (power-law) 
$M(<j)$ profiles than low-spin systems, but the scatter about this trend
is large.

Does the angular-momentum distribution correlate with the 
mass-density distribution? 
We know that the global spin characterized by $\lambda'$ does not
correlate strongly with the halo mass (Barnes \& Efstathiou 1987)
and correlates only weakly with the concentration parameter (\fig{lt_vs_c}).
Figure~\ref{fig:cm} shows $\mu - 1$ versus these mass parameters.
We detect a marginal anti-correlation with mass, with $r \simeq -0.142$ 
($p \simeq 4\times 10^{-3}$ for no correlation), and an even less significant
anti-correlation with $\cvir$, of $r \simeq -0.055$ ($p \simeq 0.26$
for no correlation).

\section{Spatial distribution of angular momentum}
\label{sec:spatial-dist}

In the previous section we explored 
the distribution by mass of the specific angular momentum within each
halo. 
This does not tell us much 
about how well the angular momentum is aligned throughout the halo
or how angular momentum is distributed spatially.  This section
is explicitly devoted to these issues. 
In \se{alignment} we address the question of alignment, 
in \se{cyl} we investigate the cylindrical symmetry of the 
angular-momentum distribution, and in \se{sph_prof} we explore 
the angular-momentum profile in spherical shells.

\subsection{Alignment}
\label{sec:alignment}

Here we address the question of how well the 
smoothed angular momentum is aligned throughout the halo volume.

\begin{inlinefigure}       
\resizebox{\textwidth}{!}{\includegraphics{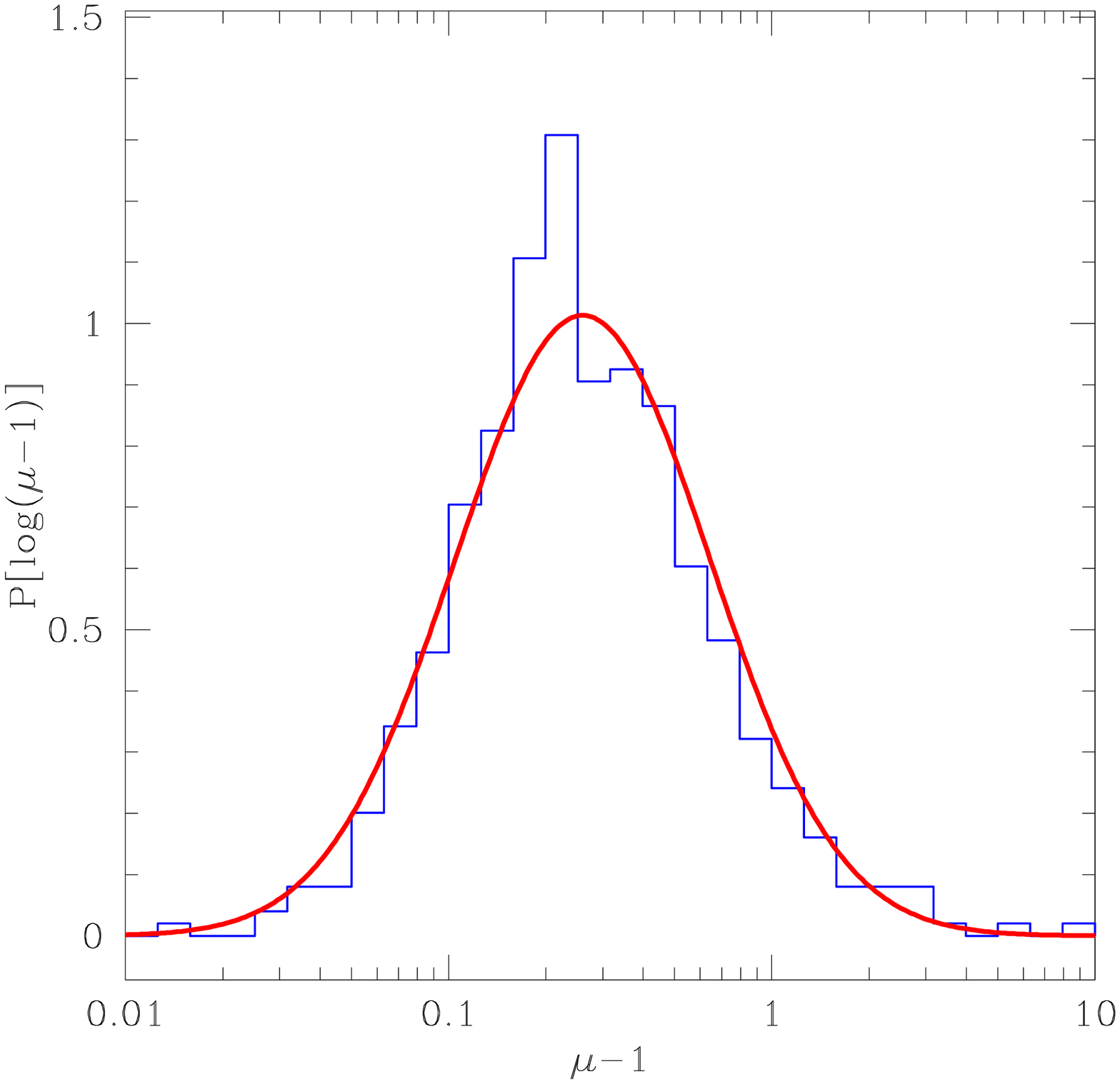}}  
\figcaption{
The distribution of $\mu$ values (histogram).  Shown is a
Gaussian distribution in $\log(\mu-1)$ (smooth curve) with
the same mean ($-0.6$) and standard deviation ($0.4$) as the
measured distribution.
    \label{fig:muhist}}
\end{inlinefigure}  

For our first alignment statistic, we divide the particles in each
halo into an inner half-mass sphere and an outer half-mass shell
and measure the cosine of the angle between inner and outer
angular-momentum vectors,  $\cos \theta_{1/2}$.  
The distribution of measured alignment values is shown in the top
panel of Figure~\ref{fig:ahist}.

Before we attempt to interpret 
these 
results, it is essential to keep in mind the large 
errors involved in determining angular momentum
direction vectors.  Even for a halo with an intrinsically
well-aligned angular momentum distribution ($\cos \theta_{1/2} \simeq 1$),
measurement errors in each of the six directional
components involved in the alignment cosine will
tend to drive the measured value {\it towards zero}, and, if
large enough, will lead
to a false measure of misalignment.  
The distribution in the top panel can therefore serve as a lower bound 
for the true alignment. 

In order to obtain a conservative estimate of the distribution of
true 
alignment cosines, we use \equ{er} to assign 
errors to each component of ${\bf J}$ in both the inner and outer
half-mass regions and perform a standard propagation of
errors to obtain 
an estimate for the measurement error on  $\cos \theta_{1/2}$.
We add the estimated error in each case to the measured 
alignment value in order to obtain a ``corrected'' distribution for
$\cos \theta_{1/2}$.  This is shown 
in the lower panel of Figure~\ref{fig:ahist}.  As argued above, we believe 
Equation~\ref{eq:er} is 
an overestimate for the measurement
error, so our corrected alignment cosines represent a plausible
upper limit for the alignment.  
The true intrinsic distribution probably
lies somewhere between the corrected and uncorrected distributions.
If we make this assumption, then the two distributions allow us to conclude
that between $70\%$ and $90\%$ of the halos are aligned
to a greater degree than $\cos \theta_{1/2} = 0.7$, although
there is also a tail of significantly misaligned halos.

\begin{inlinefigure} 
\resizebox{\textwidth}{!}{\includegraphics{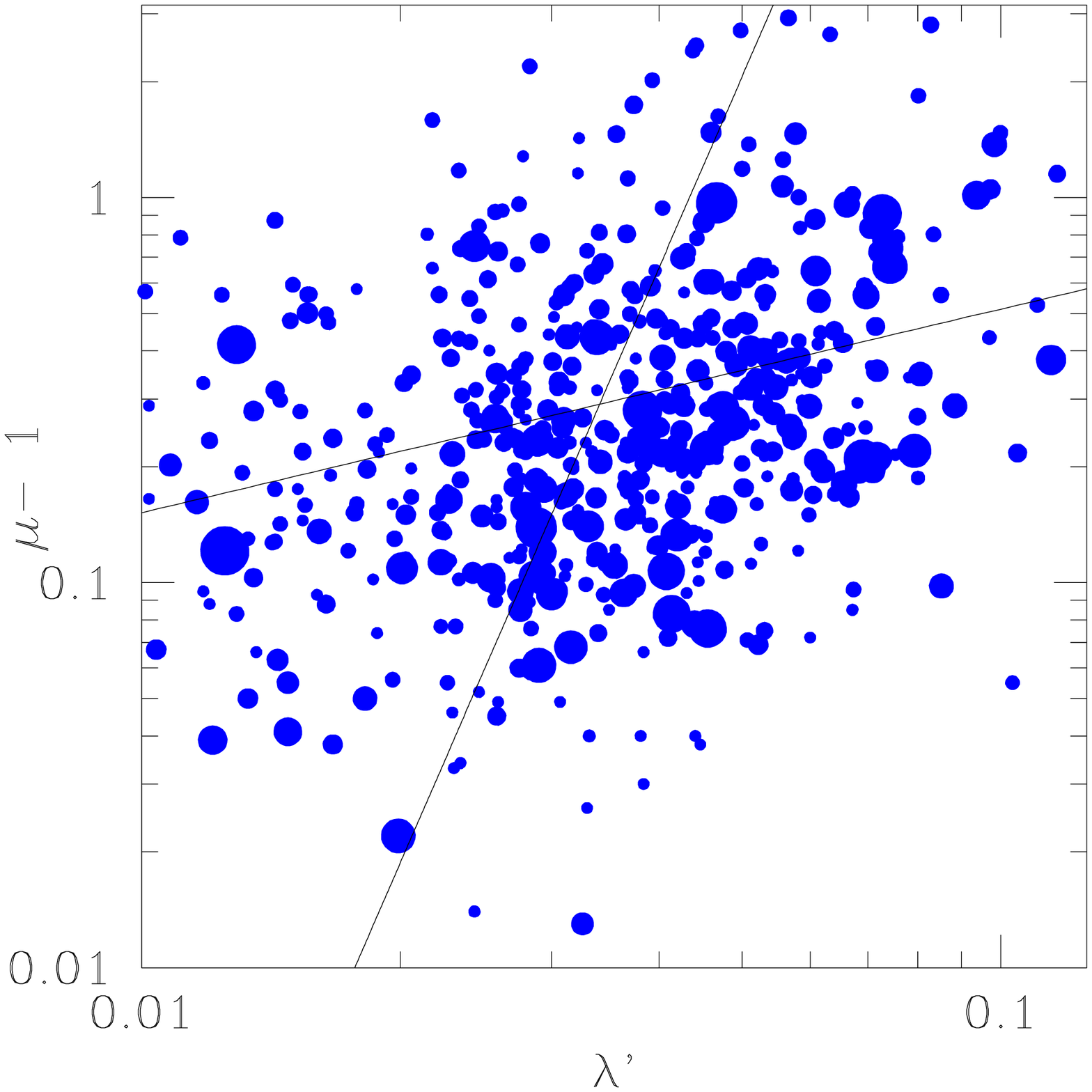}} 
\figcaption{
The joint distribution of the parameters describing the angular-momentum 
distribution, $\log(\mu -1)$ and $\log(\lt)$, showing a weak correlation. 
The symbol size is inversely proportional to the relative error on $\mu - 1$.  
The two linear regression lines are shown. 
\label{fig:mu.vs.lambda}} 
\end{inlinefigure} 

If the angular momentum within a halo is poorly aligned, one
might expect a reflection of this to appear in its $M(<j_z)$ profile.
There is a hint for such a trend in \fig{ahist}, where the distribution of 
$\cos \theta_{1/2}$ for all the halos is compared to the distribution 
for low $\mu$ halos of $\mu < 1.1$. The latter clearly show a higher 
fraction of misaligned halos.  
Figure~\ref{fig:avsmu} shows the corrected $\cos \theta_{1/2}$
values versus the shape parameter $\mu$ for each halo, and
indeed there is a significant trend, 
with $r \simeq 0.36$ ($p \sim 10^{-14}$). 
~\footnote{We also find a trend
with the measured value of $\cos \theta_{1/2}$ and $\lt$; however,
this seems to be due to the correlation between $J$ errors and
the spin.  The correlation is not apparent when we use
our corrected estimate for $\cos \theta_{1/2}$. }
The misaligned halos tend to be associated with low-$\mu$ profiles --- that is, 
profiles that deviate significantly from power laws, 
with a relatively larger amount of mass in the tails
of their $j_z$ distributions.  
This result suggests that 
there are common aspects to the origin of misalignment and that of 
the non-powerlaw nature of the $M(<j)$ profiles. 

Since halos with misaligned angular momentum distributions  
may be less likely to host large disk galaxies, 
this result should be kept in mind when modeling galaxy
formation within halos with low-$\mu$ angular momentum profiles.

\fig{jjz} addresses 
the angular-momentum alignment in halos in 
another way, independent of any specific spatial symmetry. 
Here we show the $M(<j)$ profiles for four different halos,
comparing the distributions of 
both $j_z$ and $|j|$ in cells throughout the halo volume.  
We see that in most cases, the 
two kinds of profiles are 
similar; for at least half the mass they differ by less than a factor of two.
This indicates 
that the angular momentum is reasonably well aligned throughout the halo.

\begin{inlinefigure} 
\resizebox{\textwidth}{!}{\includegraphics{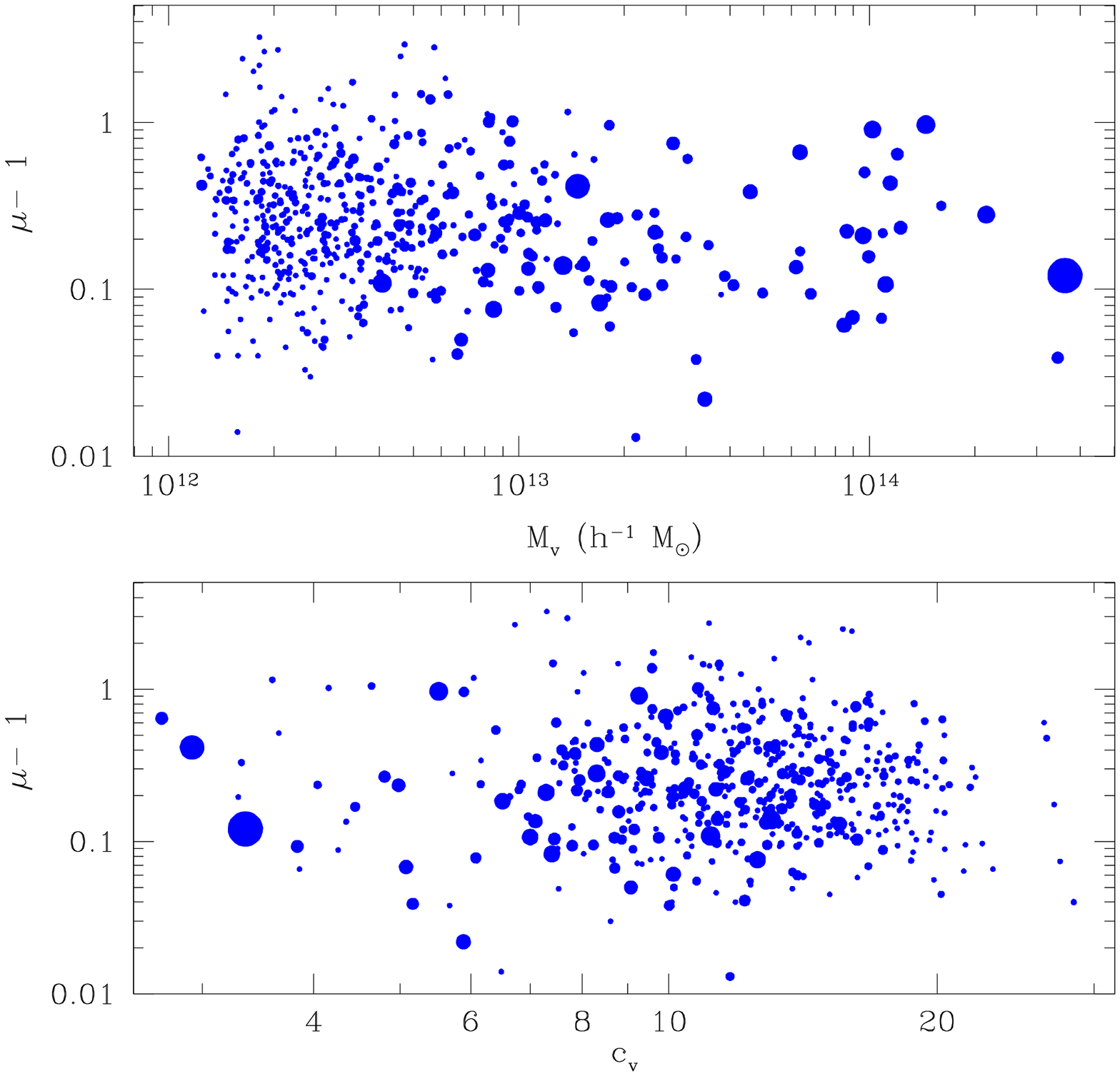}} 
\figcaption{
Angular-momentum shape parameter $\mu$ versus mass parameters.
Upper: halo mass $\mvir$.
Bottom: halo concentration parameter $\cvir$.
Symbol sizes are inversely proportional to the relative errors on $\mu - 1$.
\label{fig:cm}} 
\end{inlinefigure} 

\begin{inlinefigure} 
\resizebox{\textwidth}{!}{\includegraphics{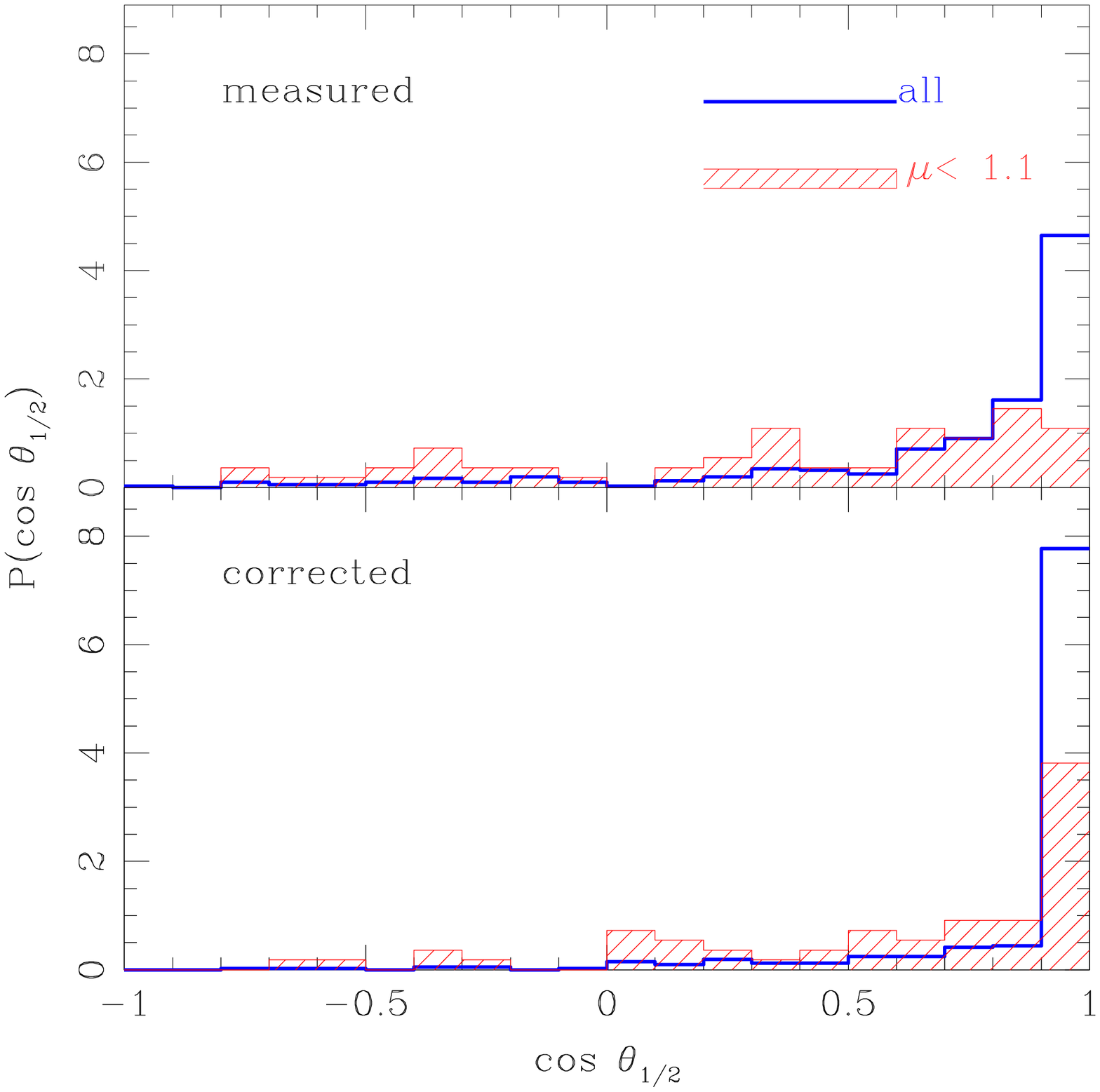}}
\figcaption{
The differential distribution of the alignment statistic 
$\cos \theta_{1/2}$,
measuring the alignment between the 
angular-momentum vectors within 
the inner and outer half mass of halos.  
Top: the measured values of $\cos \theta_{1/2}$.
Bottom:
after a conservative correction has been applied to account
for the error in direction measurement.  
The true distribution should be between the distributions shown in the 
two panels. 
The open histograms show all halos, while the shaded 
histograms correspond to halos with $\mu < 1.1$.
The histograms are normalized to yield unit integrals
over $\cos \theta_{1/2}$.
\label{fig:ahist}} 
\end{inlinefigure}

\begin{inlinefigure} 
\resizebox{\textwidth}{!}{\includegraphics{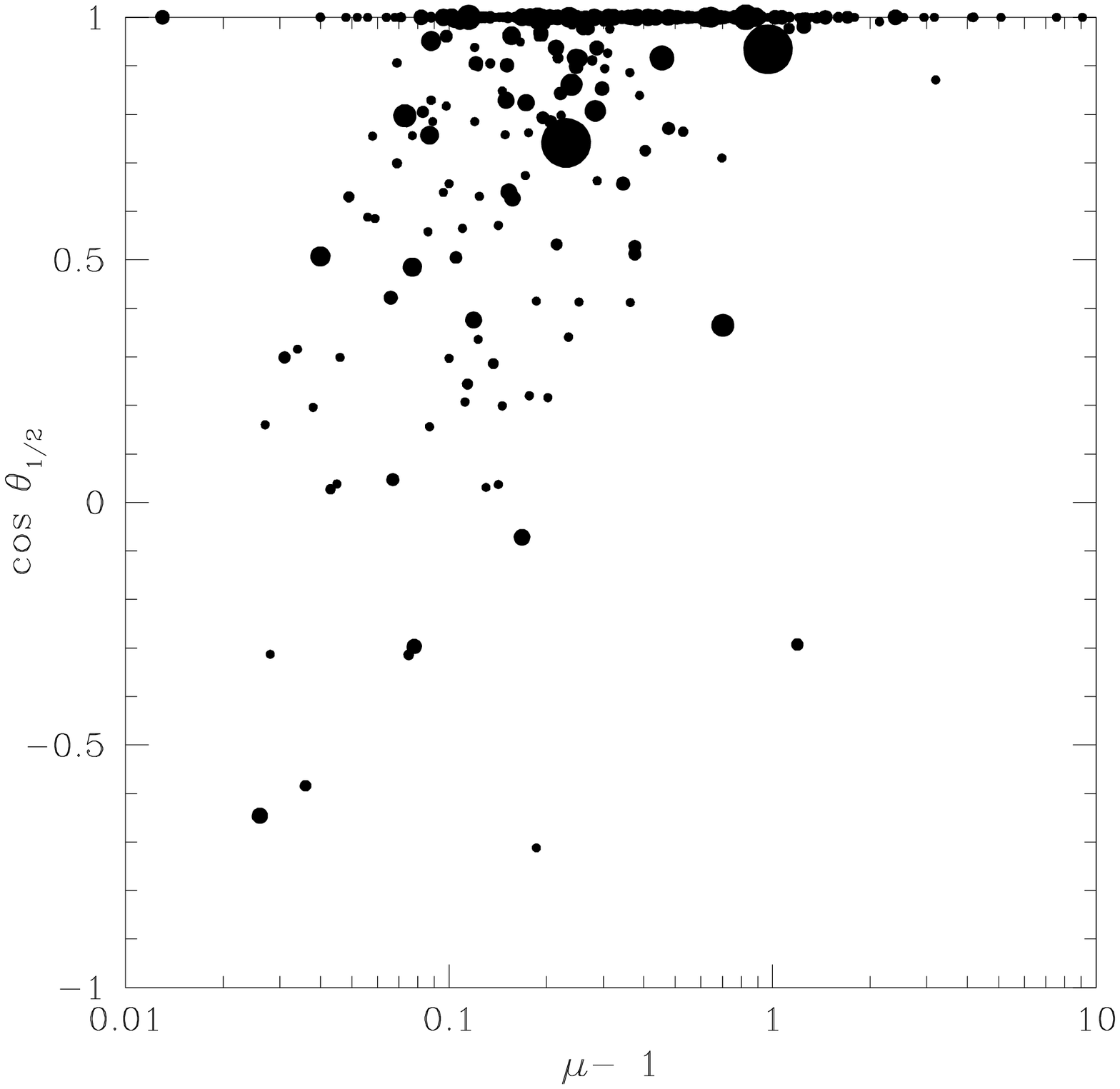}} 
\figcaption{
Alignment cosine (corrected) 
$\cos \theta_{1/2}$ versus 
angular-momentum shape parameter 
$\mu$.  
The symbol size is inversely proportional to the fit error on $\mu$.
\label{fig:avsmu}} 
\end{inlinefigure}

\vspace{1.05in}

\begin{inlinefigure} 
\resizebox{\textwidth}{!}{\includegraphics{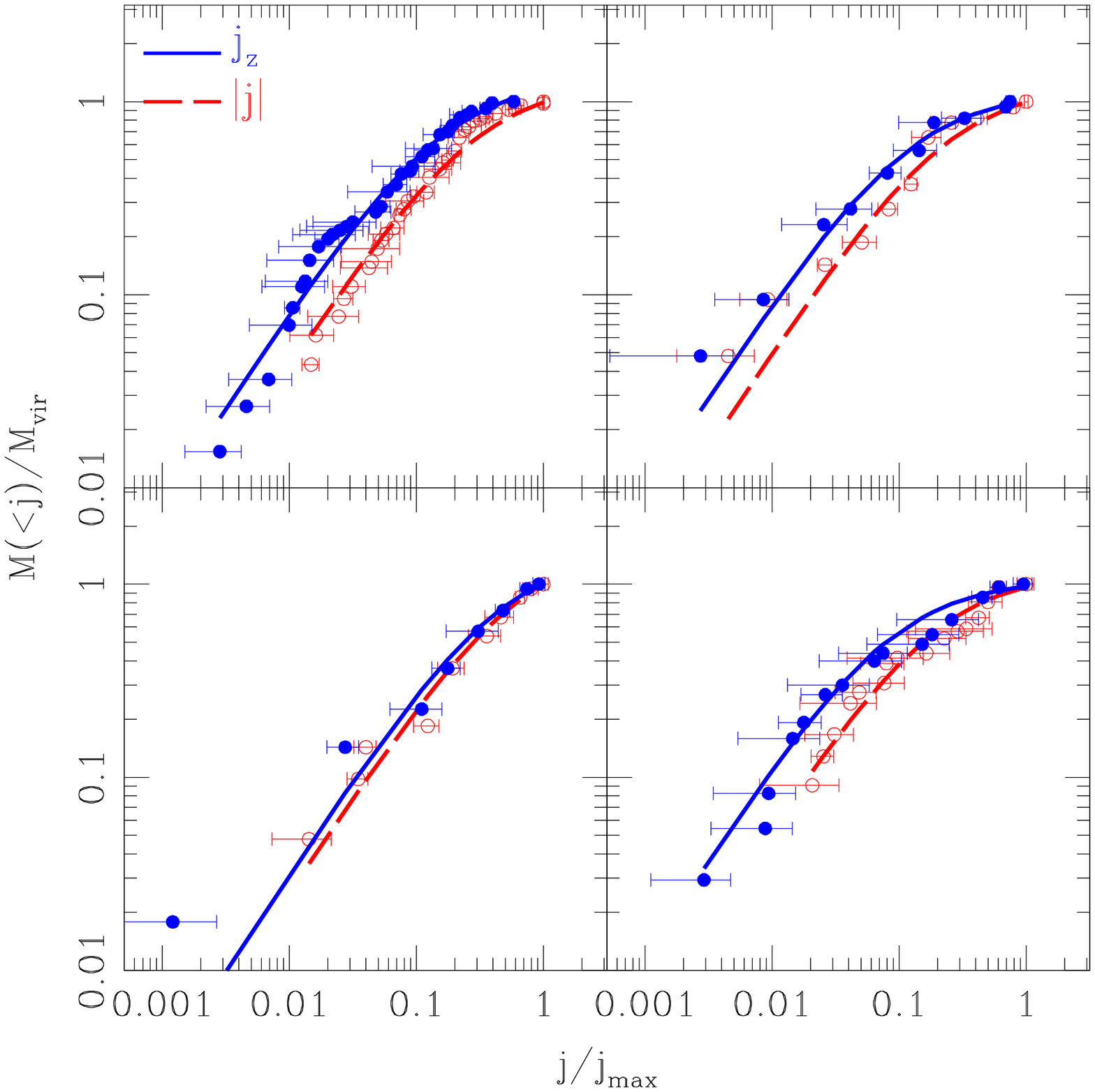}} 
\figcaption{
Alignment in four typical halos.
Shown are the profiles for the angular momentum projected in the
$z$-direction (filled points, solid line fits) 
and 
those obtained using 
$|j|$ 
(open points, dashed line fits).
    \label{fig:jjz}}
\end{inlinefigure} 

\newpage

\begin{inlinefigure} 
\resizebox{\textwidth}{!}{\includegraphics{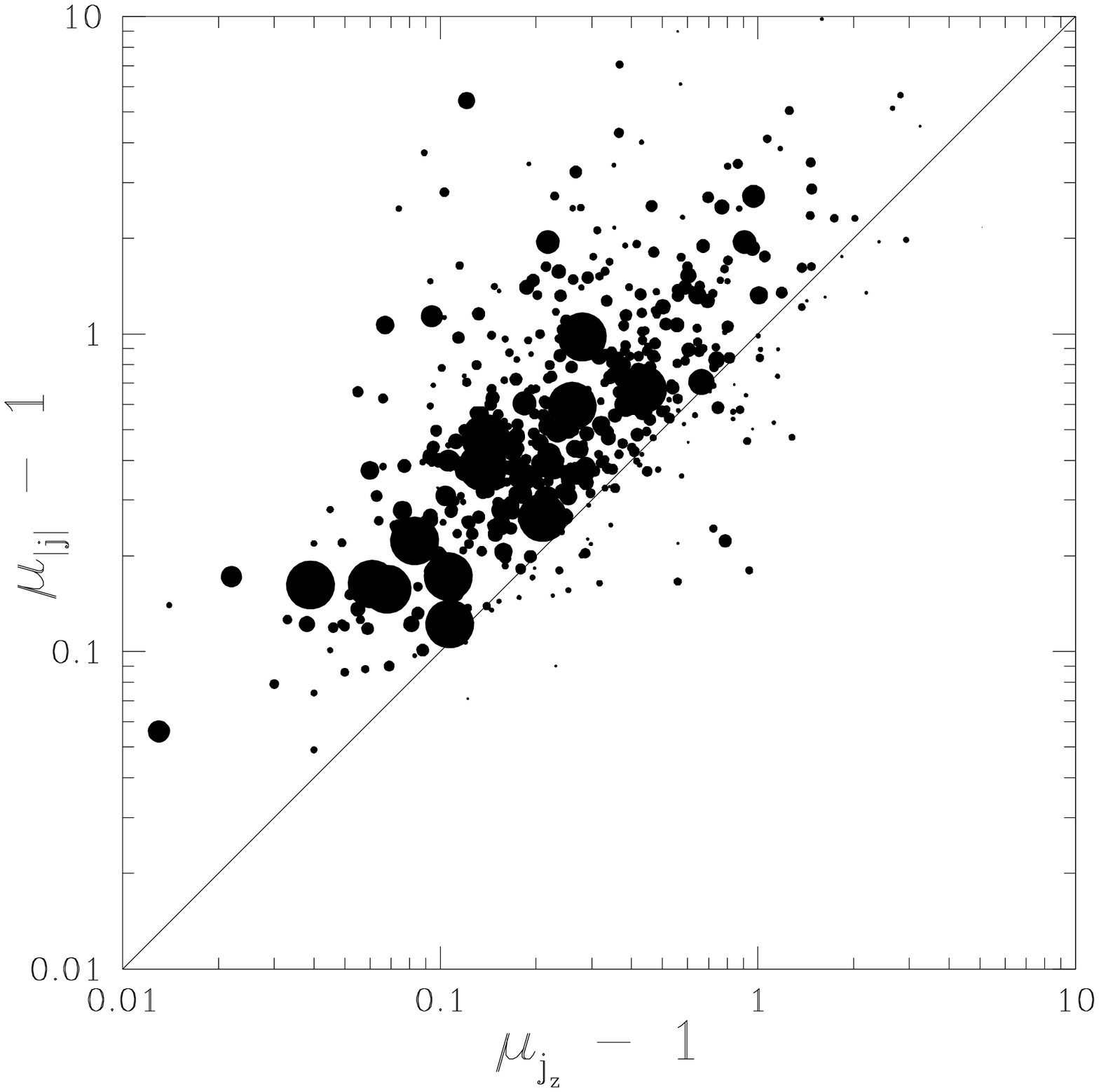}} 
\figcaption{
 The $M(<j)$ shape parameters 
for halos using $|j|$ in each cell, versus those obtained using $j_z$
in each cell.  The symbol sizes are inversely proportional to
the error on $\mu_{|j|} - 1$.
    \label{fig:mu.vs.mu}}
\end{inlinefigure} 

The slight misalignment is expressed in a 
trend that the $|j|$ distributions
systematically resemble a power-law more than the $j_z$ distributions.
This is demonstrated in \fig{mu.vs.mu}, 
which shows the 
values of $\mu - 1$ obtained using $|j|$ versus
those obtained using $j_z$.  
The correlation is obvious, with $r \simeq 0.75$. 
The value of $\mu - 1$ is typically 
about a factor of 3 higher for the $|j|$ profiles.
The fact that this factor is somewhat higher for low values of $\mu$, 
reflects the correlation seen in \fig{avsmu} between alignment and $\mu$.

\vspace{1in}

\subsection{Cylindrical symmetry}
\label{sec:cyl}

Cylindrical symmetry 
is the natural symmetry in the spatial distribution of angular momentum 
as a result of tidal-torques or a sequence of mergers.
In order to explore the 
degree 
of cylindrical symmetry,
we have again divided the halos into cells, as described
in \se{measure_prof}, except that we no
longer explicitly impose symmetry about the equatorial plane
and rotational symmetry in $\phi$.  Specifically, we define
the cells as outlined in \se{measure_prof}, but 
then, 
for each value of $r \sin \theta$, we further divide the cell particles
into four regions:
above and below the equatorial plane, and split by rotation angle
$\phi = 0 - \pi$ and $\phi = \pi - 2 \pi$.  Because each
cell now contains roughly one fourth of the particles, 
we increased our minimum halo mass threshold by a factor of four to 
$4 \times 10^{12} \hMsun$,
only 
for this section of our analysis.

\begin{figure*} 
  \centerline{\includegraphics{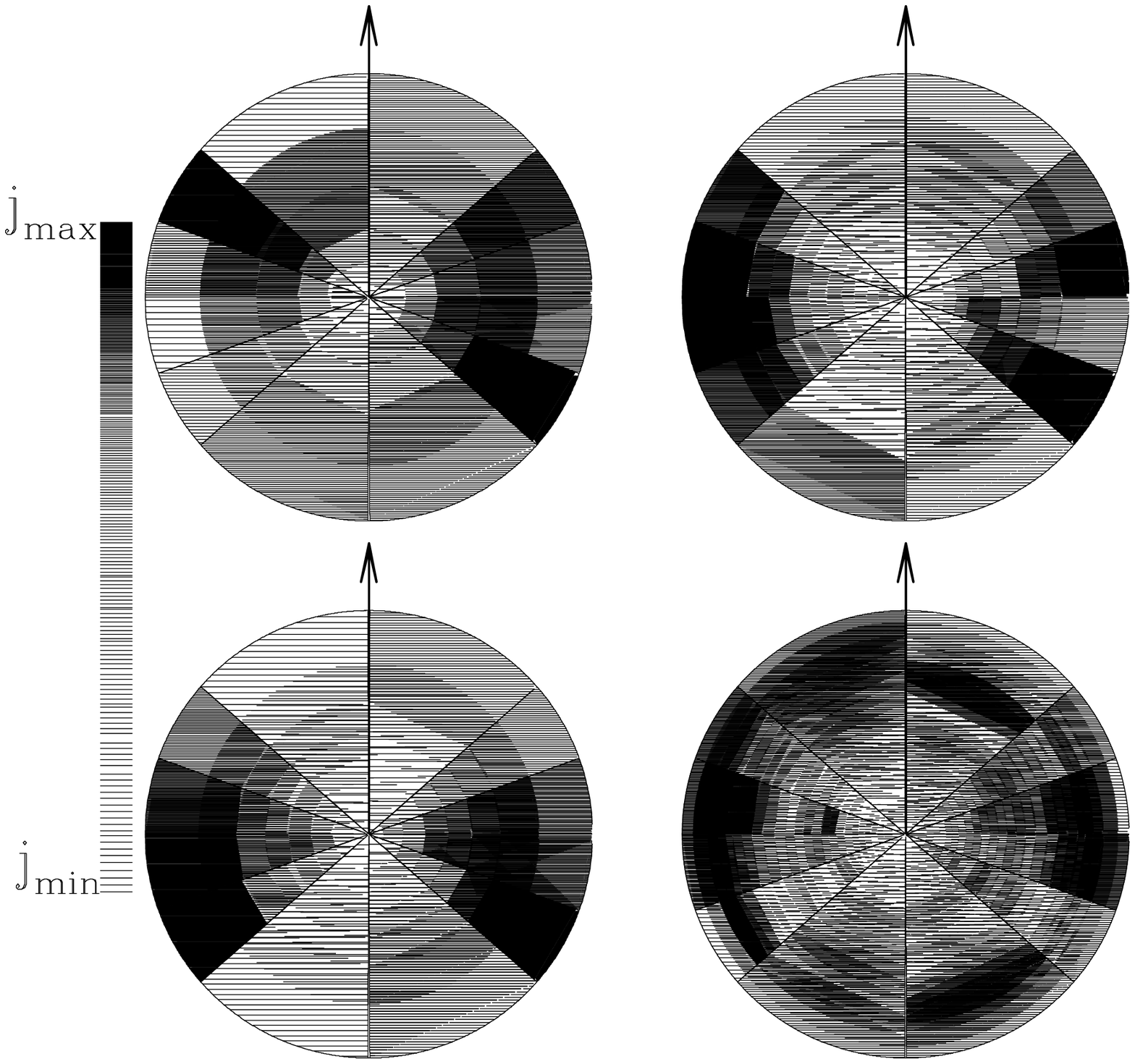}} 
  \caption{ 
Map of the angular momentum
distribution in four representative halos.  The shading code is shown, 
with the $j_{\rm max}$ cell in each halo shaded dark and the minimum
$j$ in each halo shaded light.  The arrow indicates the direction of
total $J$ in each halo. {\bf A higher quality color .eps figure is 
provided seperately on astro-ph.}
}\label{fig:map} 
\end{figure*}

Figure~\ref{fig:map} shows a map of the $j$ distribution 
in cells 
in four of our halos.
Except for the halo depicted in
the 
lower-left panel, 
the symmetry of the angular-momentum distribution tends to be more
cylindrical than spherical, with a weak dependence on $\theta$ at a constant
distance from the global spin axis, $r \sin \theta$.
On the other hand, 
the angular momentum of the halo in the lower-left panel tends to 
be concentrated 
in the $z=0$ plane.

Another way of 
visualizing the degree of 
cylindrical symmetry of halos is shown
in Figure~\ref{fig:cyl}. It shows the 
values of $j$ in the same cells and for the 
same four halos 
displayed 
in Figure~\ref{fig:map},
now 
plotted as a function of the distance from the angular momentum axis, 
$r/\Rvir \sin \theta$,
and distinguished by different symbols according to their average 
distance from the equatorial plane, $|z/\Rvir|$.   
The 
different symbol types 
should be well-mixed 
if a halo is cylindrically symmetric.  This is the case for 
three out of the four halos shown, 
except in the lower-left panel, as expected from Figure~\ref{fig:map}.

Although the angular momentum distributions in
most of the halos 
tend to qualitatively 
show a certain degree of 
cylindrical symmetry, there is some indication that, for fixed 
$r \sin \theta$, mass near
the equatorial plane tends to have more specific angular momentum
than corresponding mass in the poles.  
In order to quantify the extent to which this is true, 
we compare the quantity $j_{z=0}/j_{\rm pole}$, defined
as the ratio of specific angular momentum in cells 
near the equatorial plane ($|z/R_v| = 0 - 0.35$)
to that in cells near the poles ($|z/R_v| = 0.65 - 1$). 
For each halo, we determine $j_{z=0}/j_{\rm pole}$
by averaging over all values of $r \sin \theta$ for
which the two $z$ ranges overlap.  For example,
in Figure~\ref{fig:cyl}, $j_{z=0}/j_{\rm pole}$ can be estimated for
each halo by comparing the average value of $j$ for the solid squares 
($|z/R_v| \simeq 0.17$) to that for the open circles ($|z/R_v| \simeq 0.83$)
at fixed values of $r \sin \theta$.
Figure~\ref{fig:jrathist} shows the histogram of this ratio for
all of our halos.  
The mean is $\langle \log_{10}(j_{z=0}/j_{\rm pole}) \rangle \simeq 0.13$, 
and the standard deviation is $0.18$.

In order to test whether our specific choice of
cell geometry would bias this measure, we also generated 500 realizations
of halos with perfect cylindrical symmetry ($j_{z=0}/j_{\rm pole} = 1$).
They each had NFW density profiles and
$M(<j)$ profiles in the form of Equation~\ref{eq:fit}.   The resulting
$j_{z=0}/j_{\rm pole}$ histogram for the artificial catalog is shown
by the shaded historgram in Figure~\ref{fig:jrathist}.  The sharp peak near
a ratio of $1$ indicates that 
the bias due to 
our cell geometry
is much smaller than the detected deviation from cylindrical symmetry. 
Recall however that the average deviation from cylindrical symmetry is small,
about 35\%.

\begin{inlinefigure} 
\resizebox{\textwidth}{!}{\includegraphics{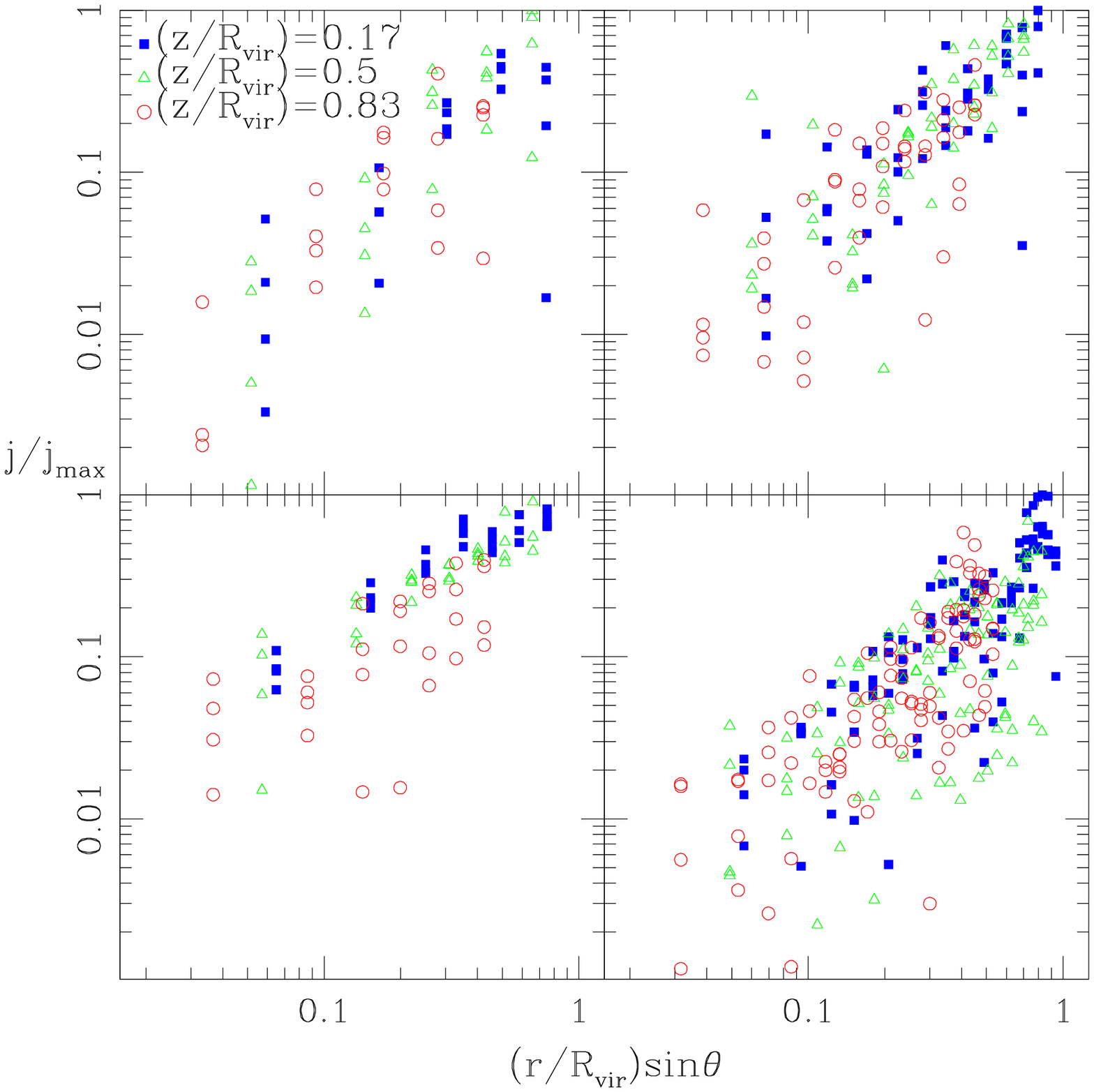}} 
\figcaption{
Cylindrical symmetry? 
Each panel shows the specific angular momentum
in spatial cells for the same halos shown in \fig{map}.
The value of $j/j_{\rm max}$ for each cell is plotted as a function
of the distance from the angular momentum axis, $r \sin \theta$.
The three 
symbol types represent the cell's average distance $z/R_{\rm vir}$
from the ($z=0$) equatorial plane,
as indicated in the figure. 
\label{fig:cyl}} 
\end{inlinefigure}

\begin{inlinefigure} 
\resizebox{\textwidth}{!}{\includegraphics{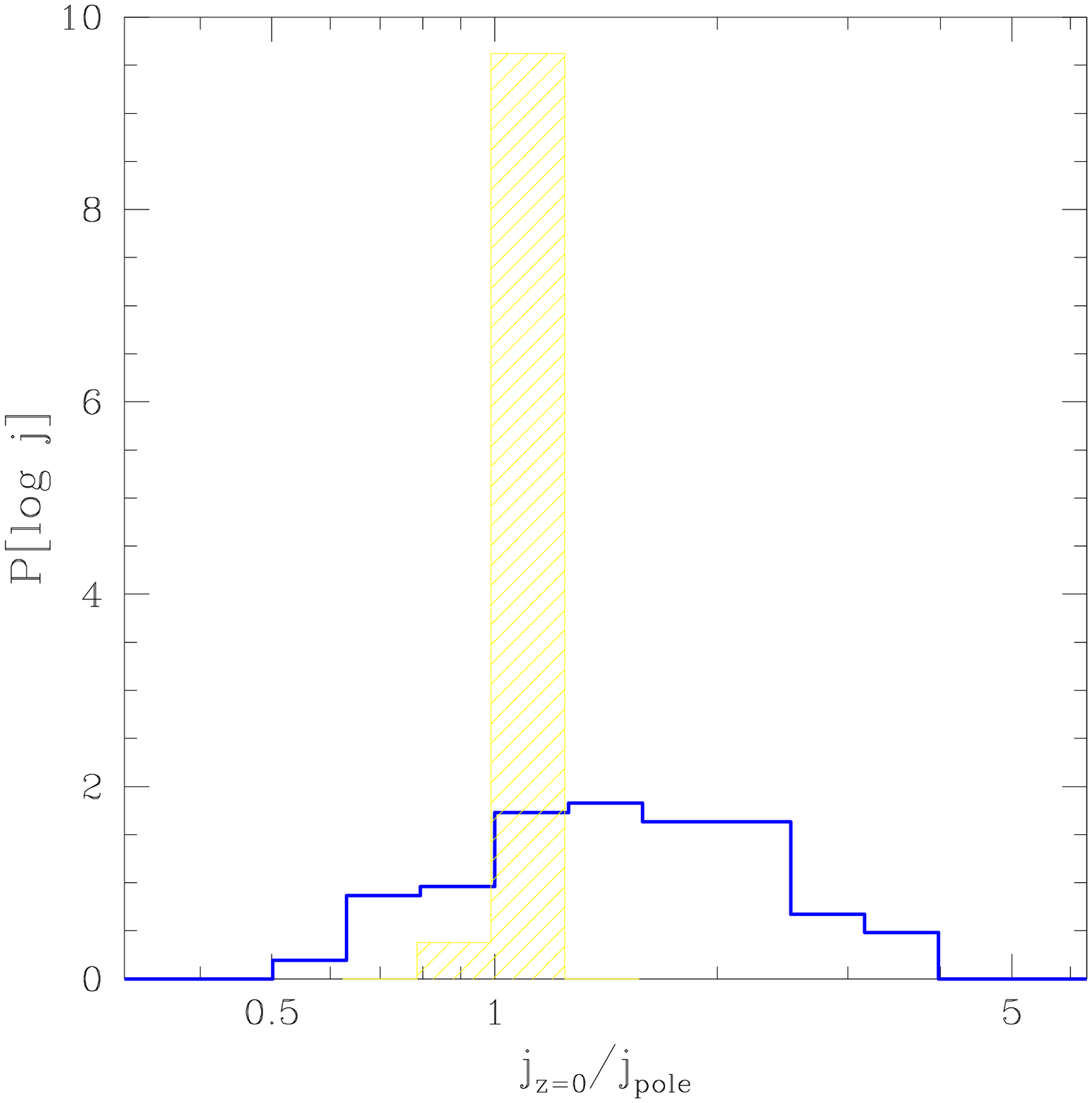}} 
\figcaption{
Deviation from cylindrical symmetry. 
Histogram of $j_{z=0}/j_{\rm pole}$ for all our halos 
(solid line), defined as the
ratio of the specific angular momentum 
about the $z=0$ equatorial plane 
and that near the poles,
averaged at fixed distances from the angular momentum axis. 
The bias due to the specific choice of cells is calibrated by the
distribution of this statistic for an artificial set of purely cylindrically
symmetric halos (shaded).  The histograms are normalized so
that they have unit integrals over $\log_{10} (j_{z=0}/j_{\rm pole})$.
\label{fig:jrathist}} 
\end{inlinefigure} 

\begin{inlinefigure} 
\resizebox{\textwidth}{!}{\includegraphics{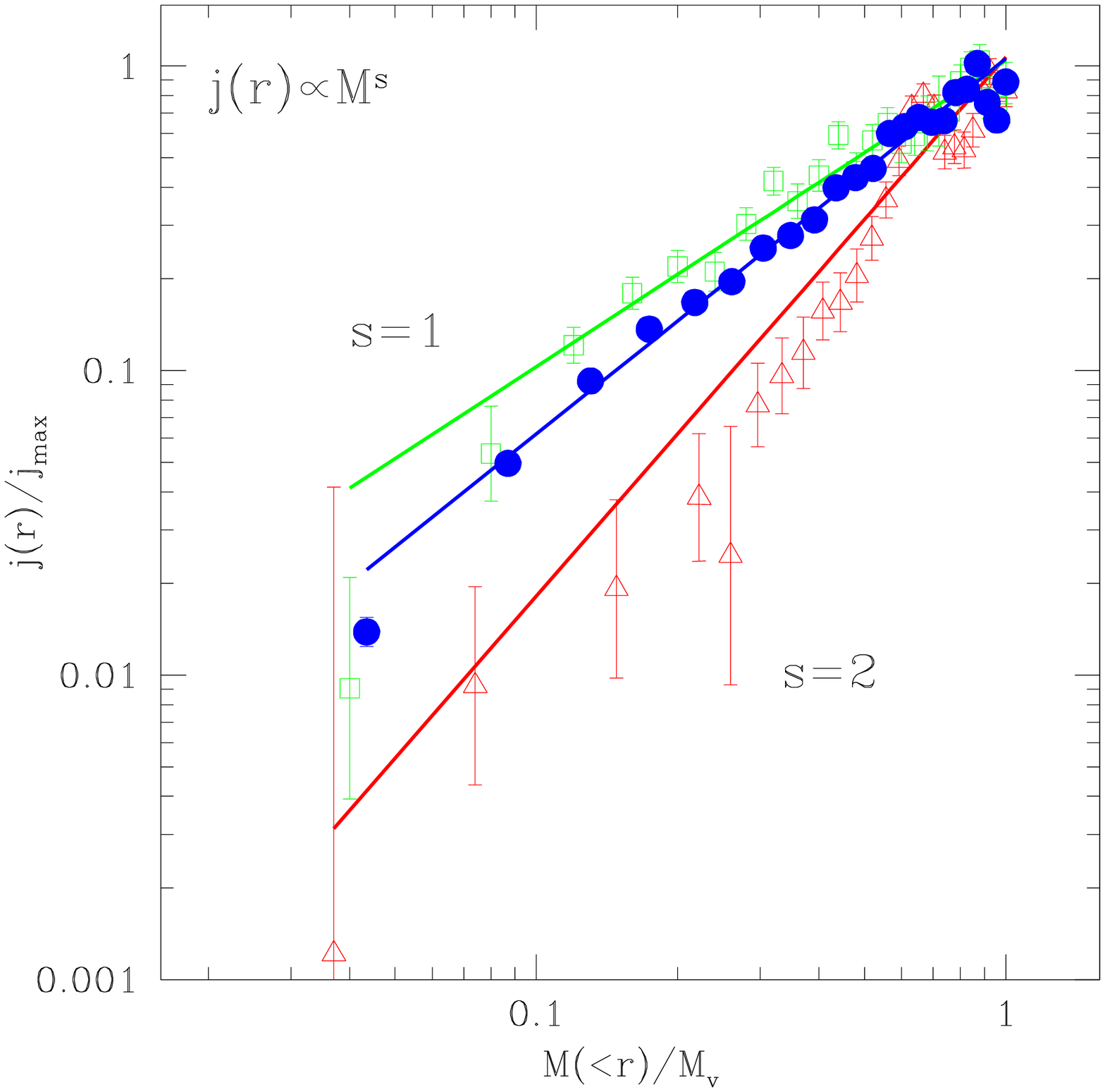}} 
\figcaption{
Specific angular-momentum profiles in spherical shells
as a function of the mass encompassed by that shell.
Shown are profiles for three halos from our simulation (symbols), 
and the corresponding power-law fits, normalized at the outermost shell.
    \label{fig:sphere}}
\end{inlinefigure}

\begin{inlinefigure} 
\resizebox{\textwidth}{!}{\includegraphics{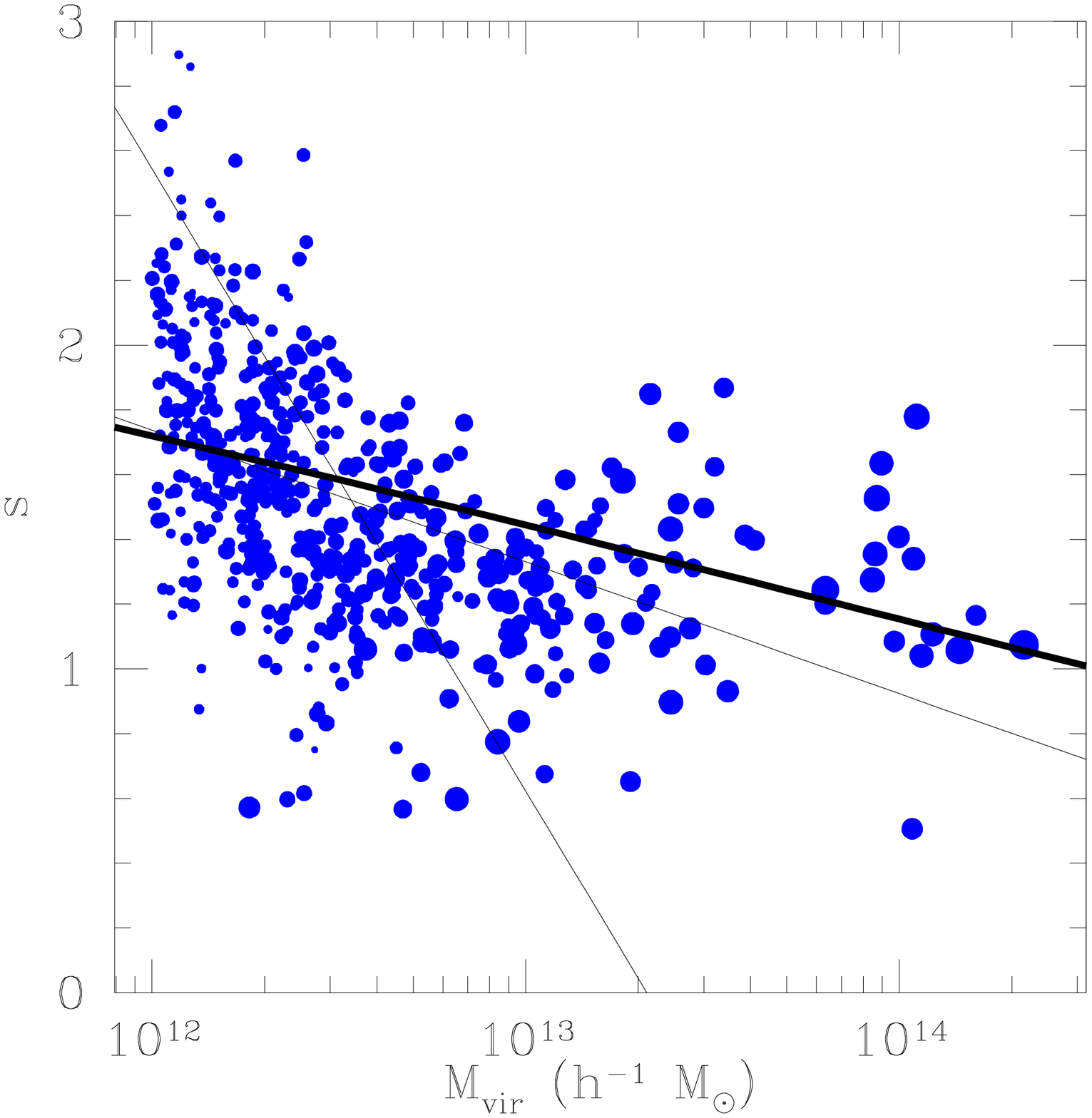}} 
\figcaption{
The slope $s$ of the angular-momentum profile in spherical shells,  
from the fit $j(r) \propto M^s$, versus halo mass. 
The symbol size is inversely proportional to the fit error on $s$.
The linear regression lines are shown (thin lines). 
Shown as a 
thick line is the prediction from linear tidal-torque theory 
(\se{origin}). 
\label{fig:s.vs.m}} 
\end{inlinefigure} 

\begin{inlinefigure} 
\resizebox{\textwidth}{!}{\includegraphics{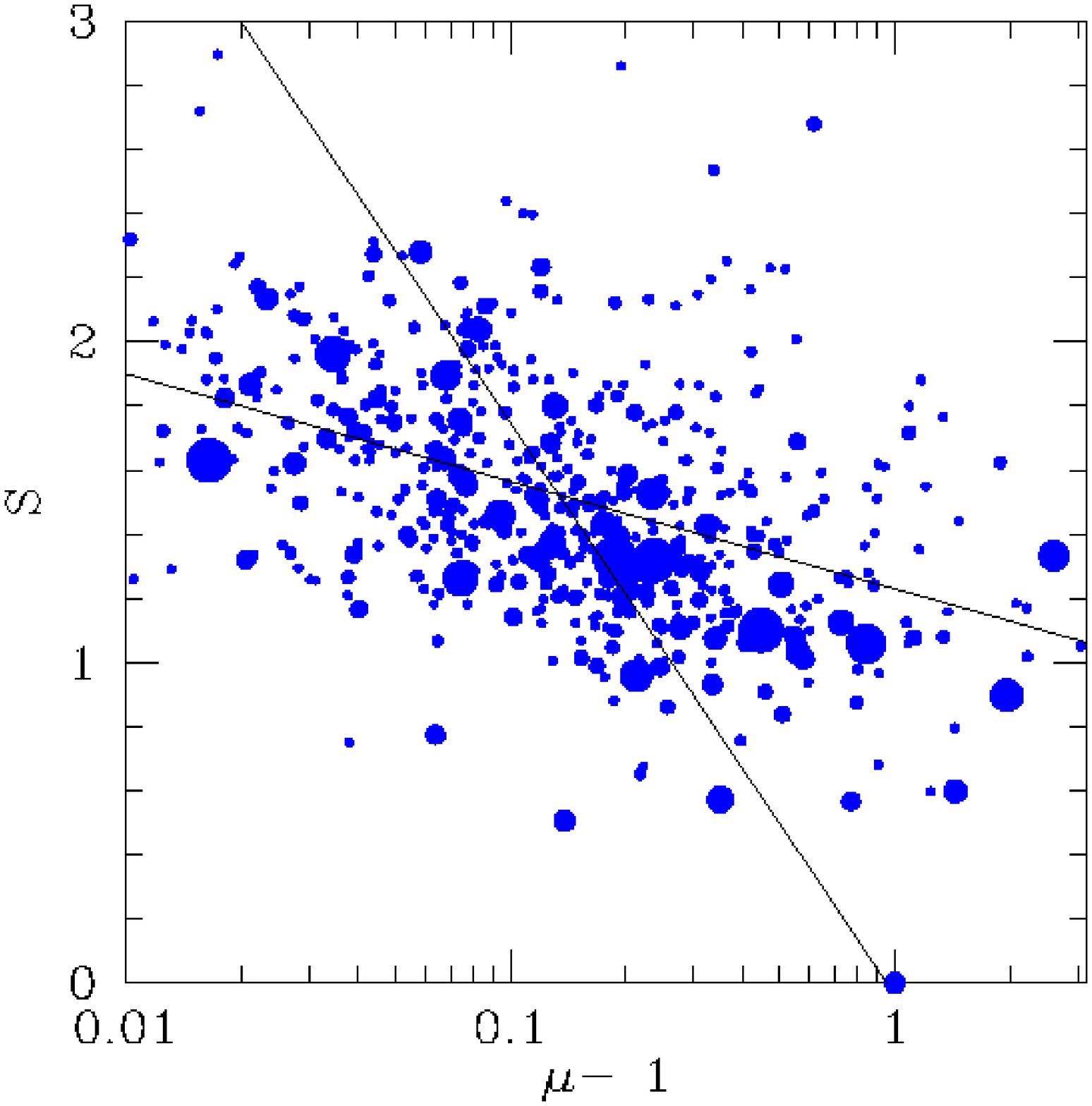}} 
\figcaption{
The slope $s$ of the angular-momentum profile in spherical shells
versus the shape parameter $\mu$ of the mass distribution of angular momentum,
$M(<j)$.
The symbol size is inversely proportional to the fit error on $\mu -1$.
Shown are the two regression lines.
\label{fig:s.vs.mu}} 
\end{inlinefigure} 

\subsection{Profile in spherical shells}
\label{sec:sph_prof}

Although, as demonstrated in \se{cyl},
the specific angular momentum distribution in halos is 
closer to being cylindrically
symmetric than spherically symmetric, an exploration of how $j$,
averaged over shells, 
behaves as a function of radius is 
of particular interest 
because this is the profile that is most directly addressed by theory 
(see below). 

Barnes \& Efstathiou (1987) found 
in their early work an indication for a rough $j(r) \propto r$ behavior. 
For our sample of well-resolved halos, we indeed
find that a power-law:
$j(r) \propto r^{\alpha}$, with $\alpha$ roughly distributed 
over the halos like a Gaussian: $\alpha = 1.1 \pm 0.3$ is
a good approximation.

In order to connect 
more closely to the mass distribution profiles discussed
in previous sections, and also to allow a more direct connection with 
theory, we focus on characterizing the spherical profile 
as a function of the mass contained within the radius $r$, $M=M(<r)$.
We  find that the power-law approximation
describes the profiles reasonably well, 
\be
j(M) \propto M^s \ , 
\ee
with $s$ roughly distributed 
over the halos like a Gaussian: $s=1.3 \pm 0.3$.

Figure~\ref{fig:sphere} shows 
three examples of the spherical $j(M)$ profiles
along with their best-fit power-laws.
Notice that although the power-laws serve as a good
general characterization of the $j$ behavior with $M$, the profiles
do not always follow a power-law form in detail. 
In many cases, the profiles show an upwards bend in the outer shells, 
reminiscent of the characteristic bend in the low-$\mu$ mass distribution,
$M(<j)$.
In some cases, $j$ is not always monotonic with $M$. 
So,  although the $s$ values provide a useful 
characteristic 
of the spherical distribution of $j$, some information is lost 
under this approximation.

Shown in \fig{s.vs.m} is the distribution of $s$ values,
versus halo mass.
Despite the large scatter, there is an anti-correlation between  
slope and mass, with $r \simeq -0.49$, ($p \sim 10^{-27}$). 
As we will discuss in \se{origin}, 
a simple 
calculation based on linear 
tidal-torque theory provides 
a hint at understanding the typical slope value 
$s \ga 1$, and predicts a mild 
anti-correlation between $s$ and the halo mass,
in agreement with the simulations. 

The spherical profile $j(M)$ and the mass distribution $M(<j)$
are expected to be related. If the mass distribution is spherically 
symmetric
and the angular-momentum distribution is cylindrically symmetric, 
this relation can be spelled out explicitly. In this case the inverse
of $M(<j)$ is the spatial profile $j_{\rm cyl}(M)$ in growing cylinders
about the total spin axis. 
A power law with a bend for $M(<j)$ translates to a corresponding power law 
with a weaker bend in the opposite direction for $j(M)$.
We thus expect a correlation between the shape parameter $\mu$ of $M(<j)$
and the slope $s$ of the spherical profile $j(M)$.
A high $s$ and a low $\mu$ both correspond to halos with
a high fraction of mass spinning slowly relative to $j_{\rm max}$.
Figure~\ref{fig:s.vs.mu} shows 
$s$ and $\mu$ for our halos.  
An anti-correlation is evident, as expected, with $r \simeq -0.36$ ($p \sim 
10^{-13}$).
The scatter reflects deviations from the global spatial symmetries. 
It is likely, 
for example, that 
partial misalignment and pockets of
low-angular momentum material at large radii contribute to the scatter.  
By comparing Figures \ref{fig:s.vs.m} and \ref{fig:cm} it is evident
that the scatter in the $s$-$\mu$ relation is 
large enough to wash away any mass dependence 
of $\mu$ that would be implied by the weak mass dependence 
of $s$.

\section{On the origin of the profile}
\label{sec:origin}

In this section we explore the possible origin of the angular
momentum profile based on two different pictures for how specific
angular momentum is acquired in halos.  In the first picture, angular
momentum is built up shell by shell, and is modeled using linear theory
with no 
further angular momentum transfer 
within the halo. 
In the second picture, the angular
momentum profile arises solely from angular momentum transfer
from one or more satellite halo merger events.  Since halo mass
accretion histories typically reflect some combination of relatively
quiescent mass accretion as well as more pronounced mergers (where $j$
transfer is likely), the
$j$ profiles of individual halos may reflect a complicated combination of
the two processes explored here. Indeed the two processes are
intrinsically linked, since, in the hierarchical framework, all
mass accretion can be treated as the accretion of smaller objects.
 As shown below, the profiles calculated
from linear theory as well as those calculated using
 $j$ transfer
from mergers produce a range of $\mu$ values, in qualitative
agreement with what is observed in the simulations.

\subsection{TTT and EPS}

Linear tidal-torque theory (TTT) provides a hint for the origin of the
detected power-law-like $j$ profile, as follows.
TTT (Doroshkevich 1970; White 1984) 
implies that the angular momentum gained by a halo
at time $t$ before its turn-around is
\be
J_i(t) = a(t)^2 \dot D(t)\, \epsilon_{ijk}\, T_{jl}\, I_{lk} \ ,
\label{eq:ttt}
\ee
where the time growth is from some fiducial initial time,
$a(t)$ is the expansion factor at $t$, 
$D(t)$ is the linear growth factor,
$I_{lk}$ is the inertia tensor of the proto-halo at the initial time
and $T_{jl}$ is the tidal tensor at the halo center at the initial time, 
smoothed on the halo scale. 
This is based on assuming the Zel'dovich approximation
for the velocities inside the proto-halo, and a 2$^{\rm nd}$-order
Taylor expansion of the potential. 
We show elsewhere (Porciani, Dekel \& Hoffman 2000)
that the standard scaling relation of TTT should be slightly modified; 
it should read
\be
j \propto D(t_{\rm c})^{3/2}\, \sigma(M)\, M^{2/3},
\label{eq:scaling}
\ee
where 
$j$ is the specific angular momentum of the halo,
$t_{\rm c}$ is the turn-around time of the halo,
and $\sigma(M)$ is the rms density fluctuation on scale $M$ at the initial
time.  
If the perturbation turned around while the cosmology was still 
Einstein-de Sitter, then $j \propto t_{\rm c}$.~\footnote{  
The $j$ dependence on the time of collapse is $\propto t_{\rm c}$
rather than $\propto t_{\rm c}^{1/3}$, because the tidal
part of the deformation tensor, the source of angular momentum,
is actually independent of the initial density of the perturbation,
which determines the collapse time. The shear depends instead on the rms
fluctuation $\sigma$, which involves an implicit mass dependence.}

To understand the general power-law behavior of the angular-momentum
profile we apply \equ{scaling}.  We can assume that mass is accreted in
shells and that the turn-around time of each shell is determined by
$\bar\delta(M) D(t_{\rm c}) \sim 1$, where $\bar\delta(M)$ is the mean
density inside $M$. For a Gaussian field, the typical density fluctuation
profile
about a random point scales like (Dekel 1981)
$ \delta(r) \propto \xi(r)$,
where $\xi(r)$ is the linear two-point correlation function.
This is accurate to a few percent also around a high peak (Bardeen
et al.~1996, Fig. 8). 
Thus, for a power-law power spectrum $P_k \propto k^n$ and a flat
universe,
we obtain the following spin profile within each halo:
\be
j(M) \propto M^{2/3 + (3+n)/3} \ .
\ee
This implies $1<s<4/3$ for $-2<n<-1$, the range appropriate for
$10^{12}-10^{14} h^{-1} M_\odot$ 
halos in a CDM spectrum, in pleasant agreement with
our finding for the simulated halos (\fig{s.vs.m}). 

The tidal torque theory can be combined with the Extended Press-Schechter
formalism (EPS, Bond et al.~1991; Lacey \& Cole 1993) to obtain a
more detailed model of $j$-distribution, as well as quantitative
predictions for values of $s$ and $\mu$ and their mass dependence.

Let us assume for simplicity that halo mass is acquired via accretion of 
material with specific angular momentum given by
Equation~\ref{eq:scaling} and that 
the 
direction of angular momentum of each
mass shell is perfectly aligned. 

We can readily estimate 
the slope $s$ of the $j(M)\propto M^s$ distribution if
we assume
that the turn-around time for each mass shell is
approximated by the time at which the mass was first accreted onto the
halo\footnote{The time of accretion is expected to be proportional to the
turnaround time; see e.g. Sugerman et al. (2000).} 
by comparing the expected $j$ value at any
two shells encompassing masses $M_1$ and $M_2$, 
\beq
s \simeq \frac{\ln[j(M_2)/j(M_1)]}{\ln(M_2/M_1)} \ .
\label{eq:slope}
\eeq
Equation~\ref{eq:scaling} provides the required value of $j$ associated 
with each mass shell given its ``accretion  time'',  $t_c(M)$. 
Considering, for example, $\Mvir$ and $\Mvir/2$, we obtain  
\beq
s(\Mvir) \simeq 1.44\, \ln \left( 2^{2/3}\, D[t_{1/2(\Mvir)}]^{-3/2}
\frac{\sigma(\Mvir)}{\sigma(\Mvir/2)} \right) \ ,
\eeq
where $t_{1/2}$ is the time at which the halo of mass $\Mvir$
first accreted half of its mass, $t_{1/2}(\Mvir) \equiv t_c(\Mvir/2)$.
We adopted for all halos, by the requirement of virialization,
$t_c(\Mvir) =t_0$ (the current age of the universe),
and set $D(t_0) =1$. The most probable value of $t_{1/2}(\Mvir)$ can be estimated via
EPS (Lacey \& Cole 1993, Eq.~2.26). The predicted $s(\Mvir)$ relation is
added to \fig{s.vs.m} as the thick solid line.  The agreement with the
mean trend shown by the simulated halos is remarkable, especially 
since our model has no free parameters.

We can further elaborate the model by following the
more detailed mass accretion history
of halos using EPS merger trees. Specifically, we have used the method of
Somerville  \& Kolatt (1999) to model mass growth and halo accretion 
histories for 
a random 
ensemble of dark matter halos formed in the $\Lambda$CDM model and
with masses in the mass range studied in the simulation. At each epoch,
$t_i$, and time interval, $\Delta t_i$, the model 
draws a 
mass $\Delta M_i$ accreted by 
a halo of mass $M_i(t_i)$ using the EPS probability
distribution. If we
assume that this mass is accreted with specific angular momentum given by
\equ{scaling} and that there is no angular momentum loss, we can
integrate the 
specific 
halo mass accretion history to get the $M(<j)$ distribution 
for the final halo at $z=0$. Figure~\ref{fig:models} shows $M(<j)$
profiles calculated using this model along with the fits of the form
$M(<j)=M_{\rm v} \mu j(j+j_0)^{-1}$,
\equ{fit},
for three representative mass accretion histories
from the ensemble of realizations.
One can see that the model 
reproduces the general shape of the $M(<j)$ profile with values of $\mu$ 
that are similar to those of simulated halos. 

For a given halo mass, the differences in merger histories lead to scatter
in the resulting values of  $\mu$, with the range of $\mu -1 \sim 0.1 - 2.5$.
The absence of small $\mu$ ($\la 1.1$) may result 
from our 
simplified model assumption
of perfect alignment.  Indeed, the range of these modeled $\mu$ values is
similar to the range obtained for our simulated halos when $|j|$ profiles
are considered 
instead of $j_z$ (see \fig{mu.vs.mu}).

We can also reproduce both the range of values and 
the trend with mass of the slope $s$ if we force a power-law fit to the
model $j(M)$ profiles. 
The halos with low (high) $\mu$ in this model are those that
accrete large fractions
of their mass at early (late) epochs.

Interestingly, the value of $\mu$ for a given halo in this model strongly
depends on 
the 
halo's formation redshift,
$z_f$, 
defined as the redshift where the mass of the most massive progenitor was
half of the final mass for the first time 
[$\mu-1 \propto (1+z_f)^{-2}$]. 
In light of this expectation, it is somewhat surprising that we find 
no noticeable 
trend with redshift for $\mu$ parameters in halos of fixed mass (see
Appendix) but we are currently investigating whether a trend is
evident using individual halo merger histories (Wechsler et al. 2001).
If there is a trend  with formation time, it may have interesting
implications for galaxy formation.

\begin{inlinefigure} 
\resizebox{\textwidth}{!}{\includegraphics{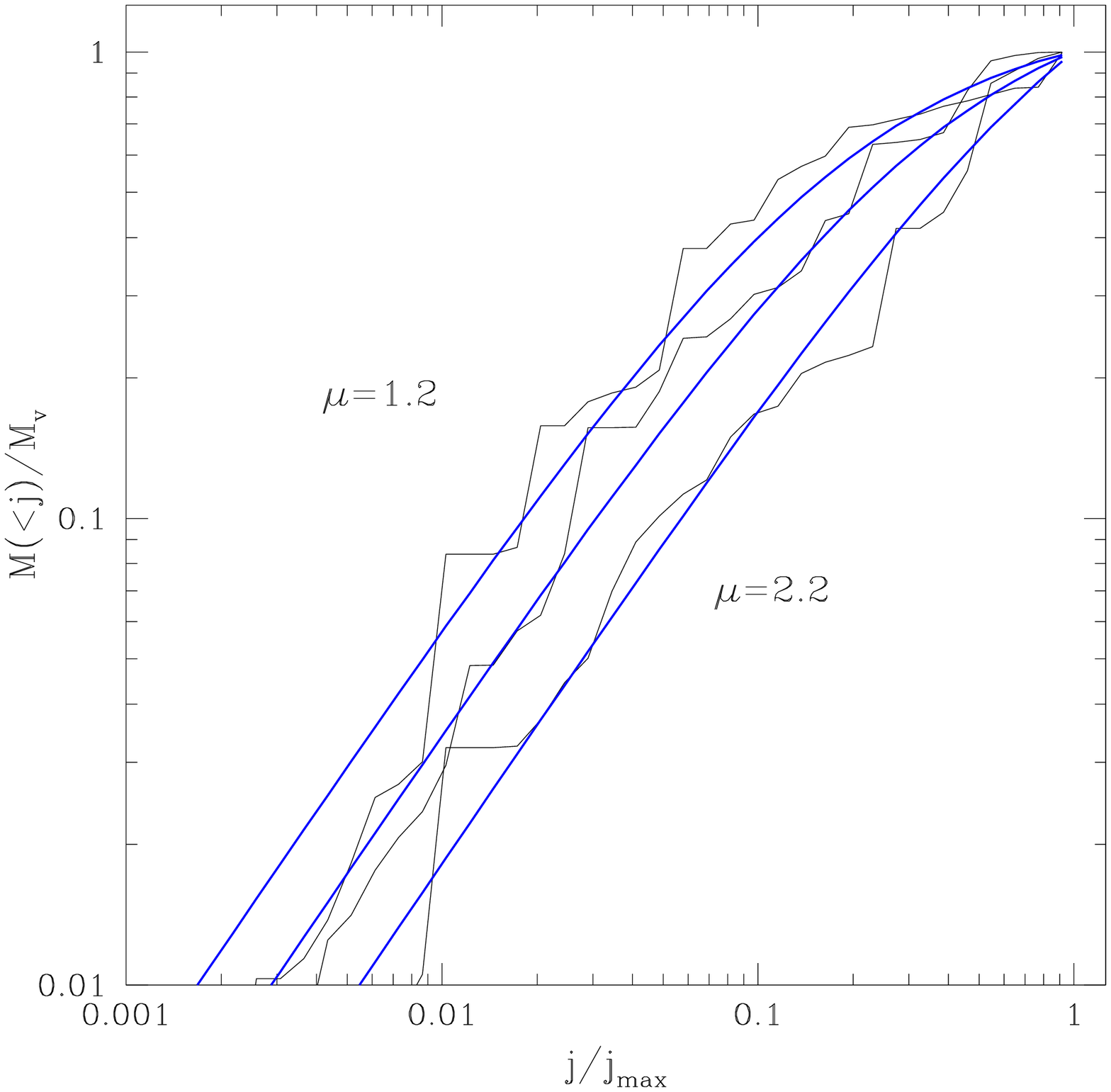}} 
\figcaption{
Angular-momentum profiles calculated using linear theory and
extended Press-Schechter merger trees.  The jagged lines are 
the derived profiles and the smooth lines are the best
fit profiles using \equ{fit}. 
\label{fig:models} } 
\end{inlinefigure} 

\subsection{Profile due to minor mergers}

A simple toy calculation for how 
angular momentum is being deposited in a large halo 
during a minor merger seems to
provide another hint for the origin of the characteristic $j$ profile.
This is studied in more detail using toy models and full $N$-body 
simulations by Dekel \& Burkert (2000).

Consider a fixed halo and an incoming satellite halo of mass profiles
$M(r)$ and $m(\ell)$ respectively.
Assume that as it moves in, the satellite 
is losing mass outside a tidal radius $\ell_{\rm t}$, 
which we approximate at $r$ to be determined by 
\be
{{m({\ell_{\rm t}})} \over {\ell_{\rm t}^2}}
={{\ell_{\rm t} \kappa (r)}\over {r^3}} \ ,
\quad
\kappa (r)=\left[ {2M(r)- r\, {\d M \over \d r}} \right] \ .
\label{eq:tidal}
\ee
For host halos and satellite halos with power-law
mass profiles
($M \propto r^\alpha, \ \alpha < 2$), 
the above approximation implies that
the total mass of a satellite halo located at a radius r of its host is
$m(r) \equiv m[\ell_t(r)] \propto M(r)$.

Suppose that the satellite is spiraling in due to dynamical friction
roughly along circular orbits,
and that the mass and $j$ are deposited locally. Then the spherical 
$j(r)$ profile is obtained by averaging over shells,
\be
4\pi r^2 \rho (r)\, j(r)
= m(r){\d [rV_{\rm c}(r)] \over \d r}
  +{\d m(r) \over \d r} rV_{\rm c}(r) \ .
  \label{eq:Jdeposit}
\ee
The two terms on the right hand side reflect 
angular-momentum transfer due to the slowdown of the satellite by
dynamical friction and the direct tidal stripping of mass from the satellite
halo, respectively.
For an isothermal host and satellite, we obtain
\be
j(M) \propto M \ .
\ee

\begin{inlinefigure} 
\resizebox{\textwidth}{!}{\includegraphics{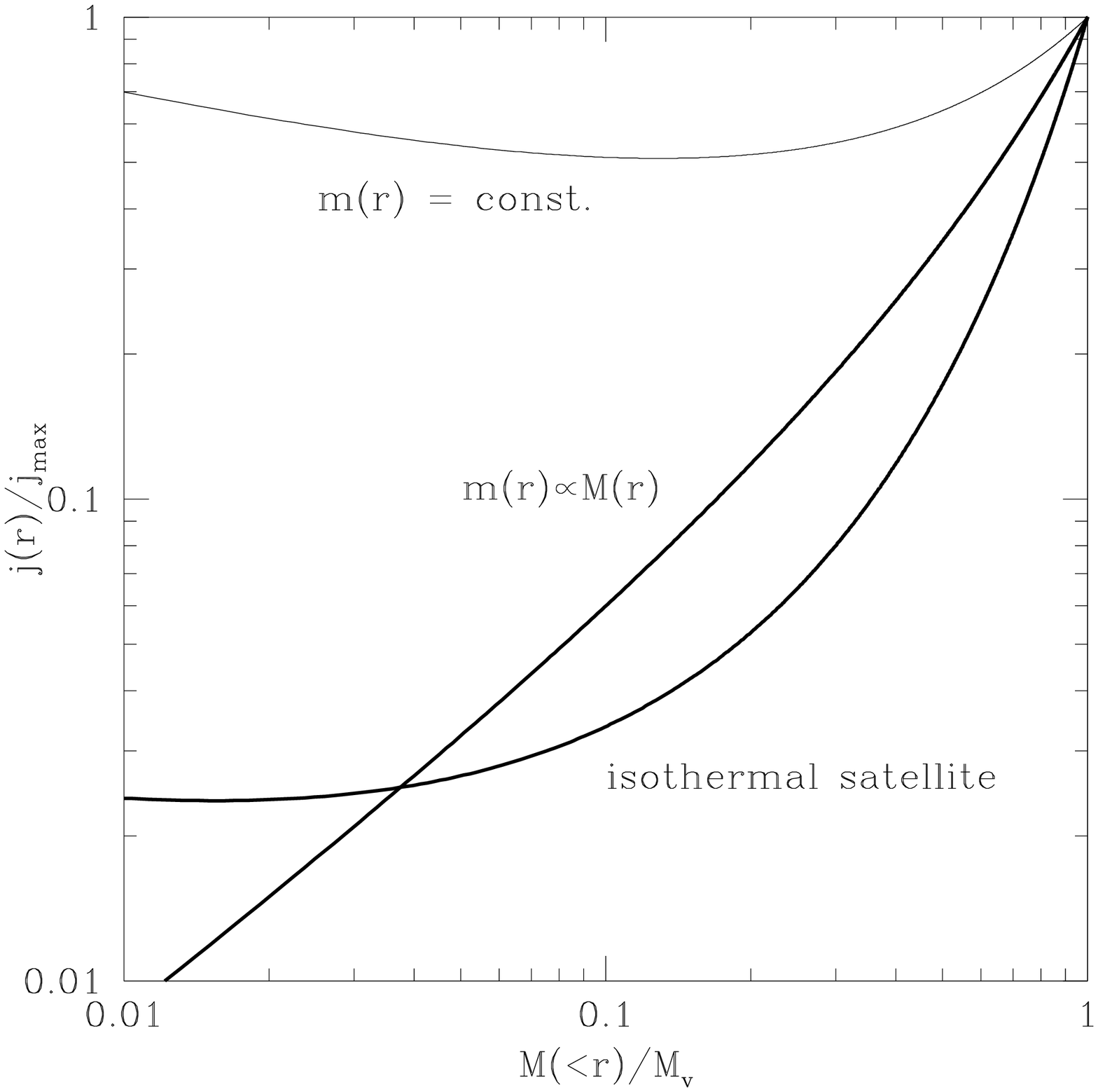}}
\figcaption{
Angular-momentum profile in spherical shells due to a minor
merger of a satellite into a fixed NFW halo, assuming circular
orbits and three different recipes for tidal mass stripping.
The marked heavy solid lines are for mass-loss recipe $m(r) \propto M(r)$
and for the mass loss in the case of an isothermal satellite profile 
respectively.
The thin solid line is for the limiting case of no mass loss.
\label{fig:fric}} 
\end{inlinefigure} 


\fig{fric} shows the expected $j(M)$ profile based on this toy model
for an NFW halo swallowing a satellite, using three alternative recipes for
mass loss by the satellite: (a) $m(r)\propto M(r)$, (b)
a satellite with an isothermal profile, and (c) a satellite of fixed mass,
for comparison. The general shape of $j(M(<r))$ in the outer decade
of mass, when realistic mass loss (a or b) 
is considered, is reminiscent of the
profiles detected in the cosmological simulation (see \fig{sphere}).

Full $N$-body simulations of mergers, spanning a range of halo and
collision parameters, confirm the robust production of such characteristic
profiles (Dekel \& Burkert 2000).
The resulting $j(M)$ from a sequence of minor mergers is expected to be
a sum of similar contributions, each projected onto the direction of
the total net angular momentum (perhaps determined by the most
major merger
involving a large fraction of the final mass).

\section{Disk Structure}
\label{sec:disk}

We now explore certain possible implications of our results on the
angular-momentum distribution in the halos for
the formation of galactic disks.   We follow the limiting case
assumption that the specific angular momentum is conserved
during gas infall into a centrifugally supported disk,  
and compare the resultant gas density profile to an exponential disk.  
Also, since the angular momentum structure of halos
is governed by at least two parameters ---  say $\mu$, in addition to $\lt$
--- we specifically explore how variations in the shape
parameter may affect the disk characteristics.

\begin{inlinefigure} 
\resizebox{\textwidth}{!}{\includegraphics{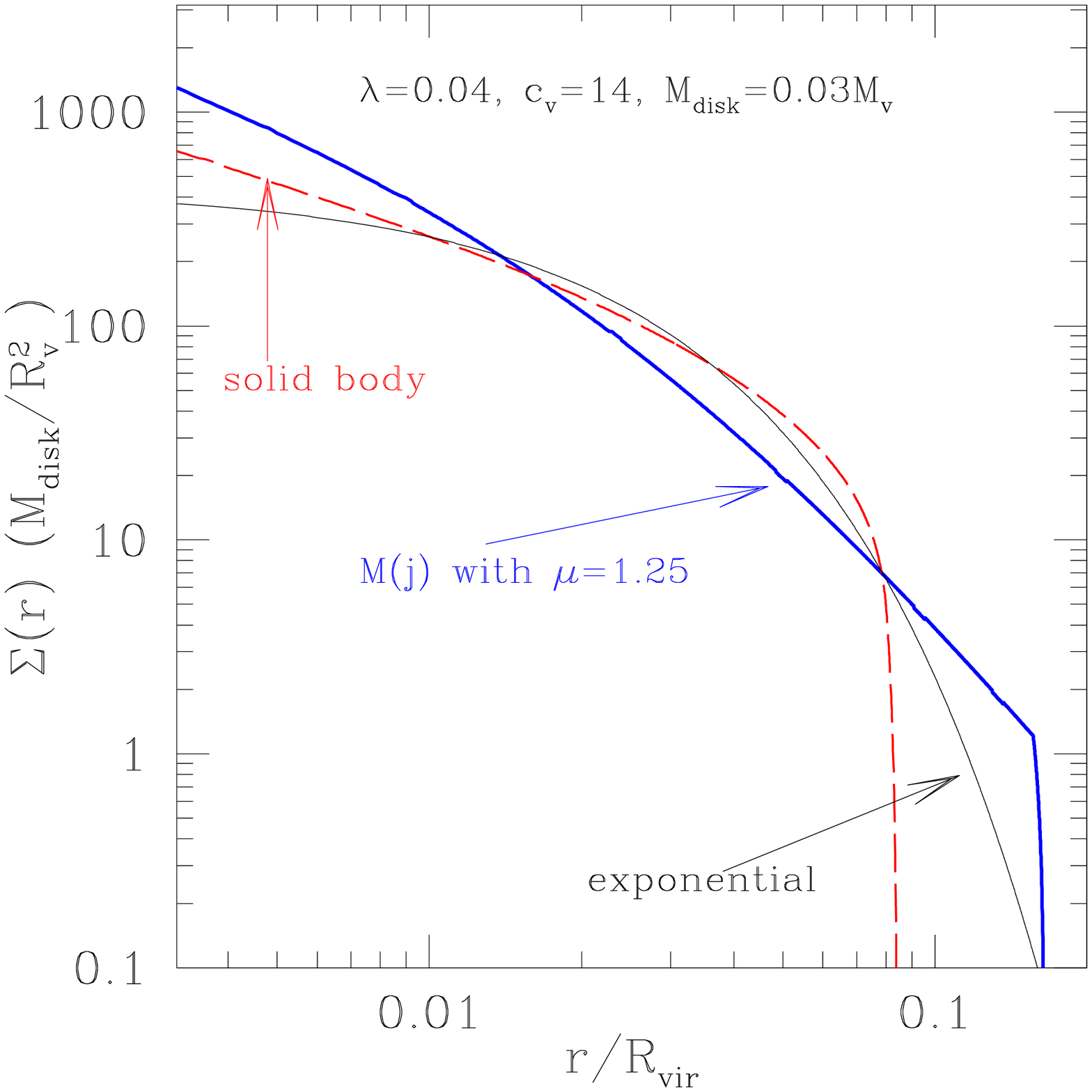}} 
\figcaption{
The 
disk surface-density profile (heavy solid line) 
implied by our $M(<j)$ distribution with the indicated parameters 
under the assumption that specific angular momentum
is conserved during adiabatic baryonic infall.  
Shown for comparison are an exponential disk
and the disk profile resulting under similar assumptions from 
a uniform, solid-body rotating sphere.
\label{fig:surf}} 
\end{inlinefigure} 

Suppose that each dark-matter halo has a specific angular
momentum profile $M(<j)$ given by \equ{fit}.
Given $\mu$, $M(<j)$ is determined by the spin parameter
via $j_0 = \sqrt{2} \lt \Vvir \Rvir b(\mu)^{-1}$.
Assume that 
initially the specific angular momentum distribution
of the gas mirrors that of the halo,
$M_{\rm gas}(<j) = f M_{\rm v}(<j)$,
where $f$ is the mass fraction of the halo that ends up as cool
baryons in the disk, $M_{\rm disk} = f M_{\rm v}$.

After infall into a disk while preserving $j$, 
the specific angular momentum of 
a gas element that 
ends up in a circular orbit of radius
$r$ is $j(r) \simeq \sqrt{G M(r) r}$, where $M(r)$ is the total
mass of DM and baryons within $r$.
All the gas with specific angular
momentum less than $j(r)$ will wind up interior to $r$.
Using \equ{fit}, we obtain for the disk mass profile
\begin{equation}
m_{\rm d}(r) \simeq f \mu \Mvir \frac{ j(r)}{j_0 + j(r)}, 
\quad j(r) < j_{\rm max}.
\end{equation}

In order to estimate the implied disk structure, and to gain a qualitative 
understanding of how the value of $\mu$ may affect the disk size and 
density profile, we assume first that the dark halo does not
react to the infall of gas and approximate the
total mass distribution of the system to be that of an isothermal
sphere, $M(r) \propto r$.  In this case
$j(r)  \simeq  r V(r) = r V_{\rm v}$.
The final mass distribution of the disk is then
\begin{equation}
m_{\rm d}(r) \simeq f \mu \Mvir  \frac{r}{r_{\rm d} + r}, 
\quad r < r_{\rm max}.
\label{eqt:mofr}
\end{equation}
Here,  $r_{\rm d} \equiv \sqrt{2} \lt R_{\rm v}b(\mu)^{-1}$
and $r_{\rm max} = r_{\rm d}/(\mu - 1)$.
The implied surface density profile is
\begin{equation}
\Sigma_{\rm d}(r) = {f\mu M_{\rm v} \over 2\pi} { r_{\rm d} \over r (r_{\rm d} +r)^2},
\quad r < r_{\rm max}.
\label{eq:sigma}
\end{equation}
In our simplified model, the surface density vanishes beyond 
$r_{\rm max}$.

\begin{inlinefigure} 
\resizebox{\textwidth}{!}{\includegraphics{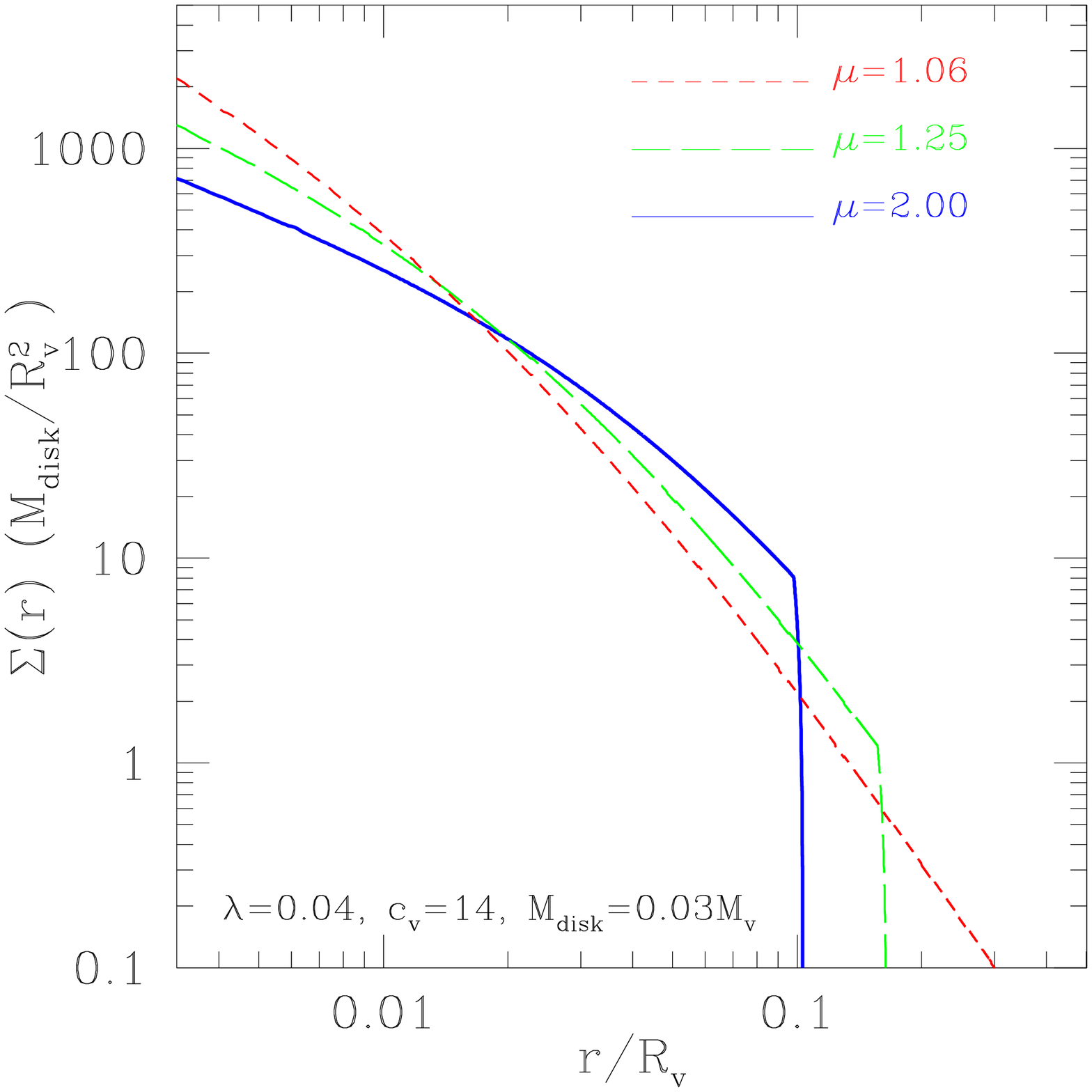}} 
\figcaption{
The resultant disk surface-density profiles, as in \fig{surf},
for three different values of $\mu$.
The abrupt edges of the disks correspond to 
$j = j_{\rm max}$ in each case.  Such an edge is clearly unphysical; 
they arise from our simplified
functional characterization of the $M(<j)$ profiles of halos.  
True profiles are likely to have smoother edges, the study
of which requires simulations with higher mass-resolution.
\label{fig:surf2}} 
\end{inlinefigure} 

Several implications are apparent from the derived disk surface-density 
profile of \equ{sigma}.  First, the distribution is qualitatively
similar to an exponential disk ---  it falls rapidly at 
large $r$, and flattens towards the origin.  
However, in detail, there are significant differences.
At large radii, $r \gg r_d$, the fall off
$\Sigma_{\rm d}(r) \propto r^{-3}$  is slower than exponential.
Near the origin, $r \ll r_d$, the profile is steeper than exponential,
$\Sigma_{\rm d}(r) \propto r^{-1}$ versus $\propto {\rm const}$.
As expected, the scale radius $r_d$ scales linearly
with $\lt$.  However, the scale radius is also 
a strongly decreasing function of $\mu$.  For a fixed spin parameter
$\lambda'$,
as $\mu$ varies over its $\sim 90\%$ range about the mean 
($\mu-1 = 0.06 - 1.0$), $r_{\rm d}$ varies by about a factor of $5$.  
This is similar to the variation of $r_{\rm d}$ as a function of $\lt$
for fixed $\mu$ (a factor of $5$ over the 90\% range of $\lt$).

Figure~\ref{fig:surf} shows the disk surface-density profile 
derived using a more accurate numerical calculation.  
Here we assume that the dark halo initially followed an NFW density 
profile with $\cvir = 14$. The $M(<j)$ profile is characterized by 
$\lt = 0.04$ and $\mu = 1.25$, and the disk mass fraction is assumed 
to be $f=0.03$.  The disk profile is obtained by self-consistently solving
for the disk circular velocity at $r$ and taking into account the
adiabatic contraction of the halo in response to the baryonic infall
(Blumenthal et al. 1986). Shown in comparison is the surface 
density profile obtained under the same assumptions regarding the
halo structure and $j$ conservation except that the given $M(<j)$ 
distribution is replaced with the assumption that the final disk profile 
is exponential.
A similar result is obtained by assuming that the initial $M(<j)$
is that of a uniform-density sphere in solid body rotation.

Figure~\ref{fig:surf2} shows how the derived disk profile varies
as a function of $\mu$.  As expected, the central concentration of the 
disk is a decreasing function of $\mu$.

As anticipated,
 compared to the exponential disk (and solid body rotator)
the derived disk profiles have core and tail excesses.
Only halos with the highest $\mu$ values ($\mu \ga 2$, corresponding
to $\sim 5\%$ of the halos) produce 
surface density profiles that begin to resemble an exponential disk.

In comparison with the observed light profiles of disks, 
the derived gas surface density profiles deviate in their general shape
and in particular in their central concentration.
The general deviation of shape is less worrisome because
the final stellar profile may not necessarily mirror the gas profile.
As Lin \& Pringle (1987) and Olivier et al. (1991) argued, 
viscous transport processes
can act to produce exponential stellar disks under 
rather general assumptions about the nature of the initial gas disk.
However, since typical viscous processes would tend to amplify the
central density even further, the indicated central density excess 
is of some concern.

This result is reminiscent of the more severe problem of
excessively small disk sizes seen in hydrodynamic cosmological simulations 
(e.g., Navarro \& Steinmetz 1997, 2000).
That we find a disk central-density excess even under the assumption
that angular momentum is conserved during the collapse
indicates that the solution to the problem may not be trivial,
because any simple angular-momentum transport mechanism would
drive the angular-momentum from the gas to the halo and from the inside out
and could only worsen the discrepancy.
The solution may require that the baryons somehow obtain an
angular momentum distribution that is biased relative to
that of the dark matter towards high-$j$. 

Alternatively, if the resultant profiles are interpreted to represent
both a bulge and a disk component, with the slowly rotating core 
associated with the bulge, then the central concentration problem
may be resolved.   Testing this hypothesis will require more
detailed modeling.  In particular, it will be necessary to match the
observed distribution of galaxy bulge-to-disk ratios as well
as disk scale lengths.   
 
Another way out may be associated with the fact that some of the approximations
adopted here may break down when the $j$ distribution is significantly
misaligned throughout the halo, which indeed preferentially happens
for low-$\mu$ halos (\se{alignment}).

\section{conclusions}
\label{sec:conc}

We have studied the angular momentum structure within
dark-matter halos in the mass range $10^{12}-10^{14} \hMsun$,
using a sample of $\sim 600$ halos in an N-body simulation of   
the $\Lambda$CDM cosmology.  
We examined in detail the more massive of these halos, 
for which the angular momentum is measured more accurately,
and then verified that the more typical galaxy-sized halos have a similar
angular momentum distribution.

Our primary result is that the mass distribution of specific angular 
momentum in halos obeys a roughly universal form, which is well fit by 
the two-parameter function $M(<j) = \mu \Mvir j/(j_0 + j)$.
For a fixed value of the spin parameter $\lt$, the parameter $\mu$ 
determines the shape of the distribution, with high-$\mu$ corresponding 
to a pure power-law and a smaller contrast between the tails of the 
distribution.  The distribution of $\mu -1$ is roughly log normal,
with the $90\%$ range spanning $\mu - 1 = 0.06- 1.0$.   
Compared to an idealized uniform sphere in solid-body rotation, the simulated 
halos tend to have more of their mass in the tails of the distribution, 
especially at small $j$.  
The shape parameter is only weakly correlated with the spin parameter,
$\mu - 1 \propto (\lt)^{\delta}$ with $\delta = 2 \pm 1.5$ and a correlation 
coefficient $r\simeq 0.23$.

Most halos have well-aligned angular momentum vectors throughout their volume.
Between 70\% and 90\% have alignment cosines greater than $0.7$
between the inner and outer half masses.  However, at least 10\% of the 
halos have significant misalignments, which may be interesting for several 
reasons.  First, they tend also to be halos with low $\mu$, indicating that 
the direction of {\bf J} in accreted material, or the halo merger history, 
plays a role in determining the shape of the $M(<j)$ profile.  In addition, 
it is unlikely that the standard simple picture of disk formation could 
be valid within a severely misaligned halo, so perhaps these objects 
typically host spheroidal stellar components. Marginally misaligned 
halos may play a role in the formation of galactic warps
(e.g., Dekel \& Schlosman 1983). 

The spatial distribution of halo angular momentum tends to be more 
cylindrically symmetric than spherically symmetric.  At a fixed distance 
from the rotation axis, mass near the equatorial plane typically has only 
about $\sim 35\%$ more specific angular momentum than corresponding mass 
near the poles.  
 
The mean $j$ is spherical shells encompassing mass $M$ is well-fit
by a power law, $j \propto M^s$.  The power $s$ is distributed like a Gaussian
with a mean of $s = 1.3$ and standard deviation $\sigma = 0.3$.
 
We 
pointed out two possible explanations 
for the origin of the universal angular-momentum profile.
The first 
is based on applying a corrected version of the linear tidal-torque theory 
(Porciani \& Dekel 2000) 
to extended Press-Schechter mass accretion histories. 
The second is based on the nonlinear process of angular momentum transfer 
from satellite orbits in halo mergers 
(see also Dekel \& Burkert 2000). 
Since halo mass growth arises as a combination of merger events and 
relatively quiescent mass accretion, the origin of the profile in 
individual halos may reflect some combination of the two processes.  
Each of these processes seems to produce 
$j$ profiles that are similar in form to those observed
in simulated halos, and thus may provide a starting point for a
deeper understanding of the origin of halo angular-momentum structure.

Finally, we have started to explore the implications of our universal $M(<j)$ 
distribution 
in the context of the archetypical model of galactic disk formation,
namely, adiabatic baryonic infall to a rotationally supported disk
while conserving angular momentum in every mass element. 
We find that the implied surface-density profiles,
which vary as a function of both $\mu$ and $\lt$,
deviate significantly from an exponential disk for all but the
largest values of $\mu$.  The resultant surface profiles are more
extended than exponential at large $r$, and are overly concentrated at
small $r$.  Since the observed light profiles of disks are closer to
exponential, our result indicates that the disk formation process
cannot be fully understood within this simplified picture.

We mentioned that 
the general deviation 
from an exponential surface density at moderate radii 
might be overcome in this picture if a viscous transport mechanism 
is included (Lin \& Pringle 1987).  As  shown by Olivier et  al. (1991), 
such  a mechanism tends to produce exponential stellar distributions 
regardless of the initial gas profile.   However, most of the expected 
$j$ transport typically acts to increase the central mass density, so 
the deviation at small radii may be a significant problem.  Even though 
we have assumed that the baryons experience no angular-momentum loss, we 
ended up with an  ``angular momentum problem'' similar to that 
detected in hydrodynamic simulations (e.g., Navarro \& Steinmetz 1997, 2000,
where it is associated with $j$ transport from the gas to the dark matter 
and from inside out).  This problem is most severe for halos with small $\mu$, 
so the tendency for these halos to have misaligned angular-momentum 
distributions, which makes them 
less-likely hosts of 
large disk galaxies, 
may partly ease the problem in the limiting case where angular-momentum 
is conserved.  However, even if we focus on halos that tend to be well aligned, 
with $\mu \ga 1.1$, the central densities remain higher than those in 
exponential disks.

A solution to this problem may be to associate the derived central 
mass concentrations with bulges.  It is indeed possible that the central 
regions of disks with profiles as in \fig{surf2} are unstable to self gravity. 
More detailed modeling is needed for testing the viability of this solution. 
In particular, one may be worried about the model over-predicting bulge 
to disk ratios and in particular under-predicting the number of bulge-less 
galaxies.  On the other hand, such an interpretation may shed new light on 
the very 
origin of the Hubble sequence, since both $\mu$ and $\lt$ may play a 
role in determining disk scale lengths, surface brightnesses, and bulge
fractions.

\section*{Acknowledgments} 
This work has been supported by grants from the US-Israel BSF, 
the Israel SF,
and  NASA and NSF at UCSC and NMSU.   
JSB was supported by NASA LTSA grant NAG5-3525 and NSF grant AST-9802568.   
Support for AVK  was provided 
by NASA through Hubble Fellowship grant HF-01121.01-99A from the Space 
Telescope Science  Institute, which is  operated by the Association of 
Universities  for Research  in  Astronomy, Inc.,  under  NASA contract 
NAS5-26555.  CP was supported by a Golda Meir
fellowship. We thank 
Andi Burkert, Stephan Corteau, Sandra Faber, Ricardo Flores,
Savvas  Koushiappas,
Ari Maller, Yair Sigad, Rachel Somerville, Frank van den Bosch, 
Maya Vitvitska, David Weinberg,  and Risa Wechsler 
for useful discussions.

\newpage

\begin{appendix}

Since our error estimates are only approximate, it is important to 
carry an independent test of the validity of our $M(<j)$ profiles  
in view of the finite mass resolution of the simulation.           
The test utilizes the output from  
another $\Lambda$CDM simulation with identical cosmological
parameters and the same number of particles as used in our
main simulation,
but now 
in a box of side $30 \hMpc$, 
namely half the original size. 
The mass resolution is thus $8$ times higher, so a statistical comparison 
with halos analyzed from our lower resolution simulation at fixed mass 
will provide a useful test for the effects of mass resolution.
The smaller box simulation also allows us to extend our analysis
down to much smaller galaxy-mass halos ($\Mvir \sim 10^{11} \hMsun$).
Unfortunately, 
the higher resolution simulation was stopped at $z=1.7$, so we 
cannot compare directly to the $z=0$ results presented in the main paper.  
However, 
a comparison performed at high $z$ can perfectly serve our purpose. 
This also allows us the opportunity to check for
any evolution in the $M(<j)$ profiles with redshift.

Figure~\ref{fig:2box}
shows the best-fit $\mu$
parameters as a function of halo mass 
for halos from our main simulation at $z=3$.
The distribution of $\mu$ values is very similar to that seen at
$z=0$ (\fig{cm}), so there is no indication of evolution.
Figure~\ref{fig:2box}
also shows $\mu$ versus $\Mvir$ for halos from the higher 
resolution simulation at the same time.
Although the overlapping region in mass is rather small,
there is no indication of any offset between the two simulations.
The halos have a similar distribution of shape
parameters over about three orders of magnitude in
mass, $\sim 10^{11} - 10^{14} \hMsun$.

We conclude that mass resolution does not seem to 
limit our analysis, and also that the $M(<j)$ profiles of halos
show very little variation 
as a function of redshift and mass.

\begin{inlinefigure} 
\resizebox{0.6\textwidth}{!}{\includegraphics{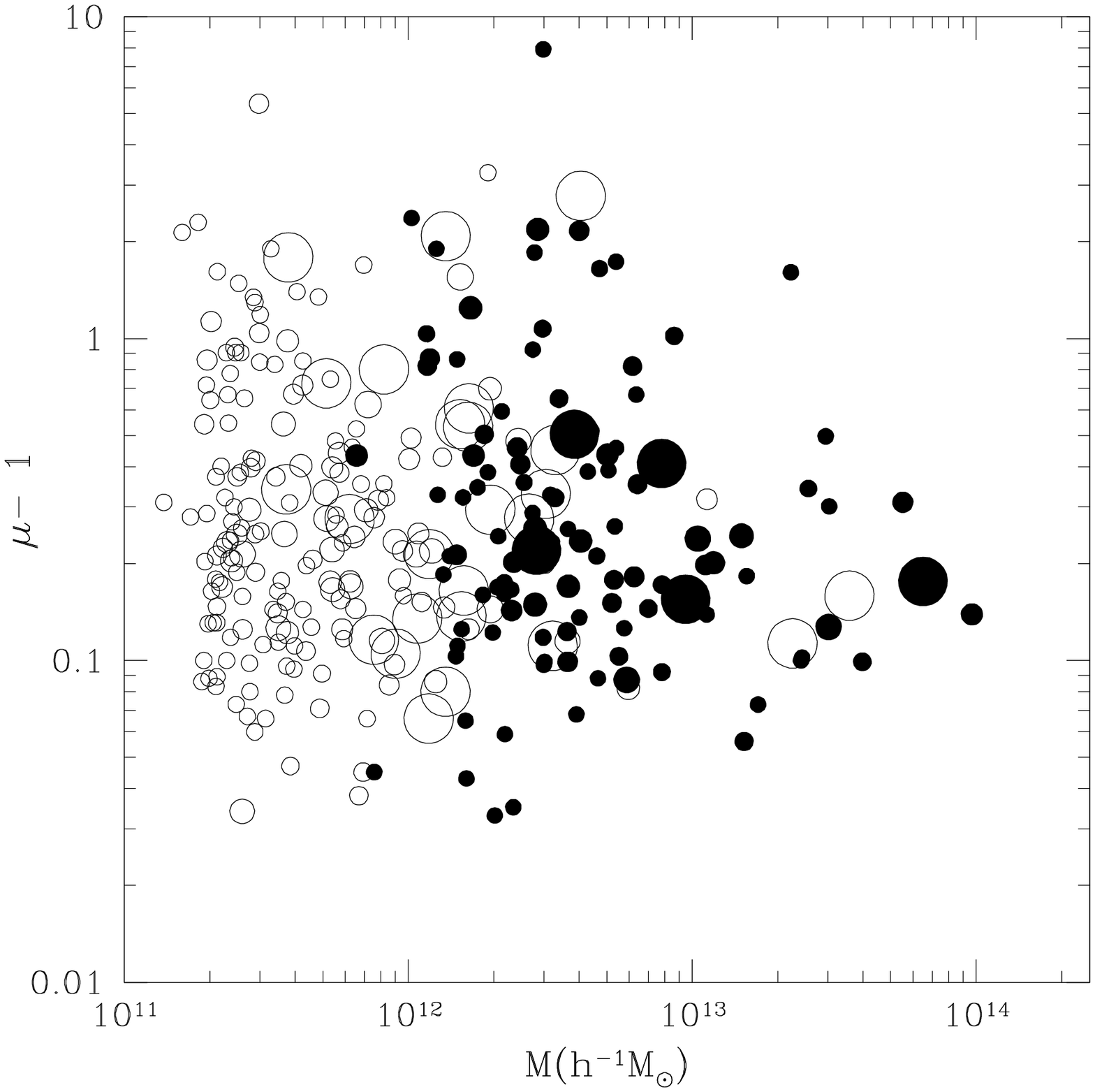}} 
\figcaption{
The effect of mass resolution: 
$\mu$ versus $\Mvir$ (at $z=3$) for 
halos in our main simulation (solid symbols) and 
in a simulation with $8$ times 
higher mass resolution (open symbols).  
The symbol size is inversely proportional to the relative
errors on $\mu - 1$.
\label{fig:2box}} 
\end{inlinefigure}

\end{appendix}

\newpage

\def\re{\reference}


\begin{references} 

\reference{bbks:86} Bardeen, J.M., Bond, J.R., Kaiser, N., \& Szalay, A. S.
1986, ApJ, 304, 15

\reference{be:87} Barnes, J., \& Efstathiou, G. 1987, ApJ, 319, 575


\reference{bcek:91} Bond, J. R., Cole, S., Efstathiou, G., \& Kaiser, N.
1991, ApJ, 379, 440

\reference{bn:98} Bryan, G.L., Norman, M.L. 1998, ApJ, 495, 80 


\reference{bffp:86} Blumenthal, G.R., Faber, S.M., Flores, R., \& Primack, J.R.
1986, 301, 27

\reference{bfpr} Blumenthal, G.R., Faber, S.M., Primack, J.R., \& Rees, M.J.
1984, Nature, 311, 527

\reference{profiles} Bullock, J.S., Kolatt, T.S., Sigad, Y., 
Somerville, R.S.,  Klypin, A.A., Primack, J.R., \& Dekel, A. 2000, 
MNRAS in press (astro-ph/9908159) 


\reference{br:85} Burstein, D, \& . Rubin, V. C. 1985, 297, 423 

\reference{ct:96} Catelan, P., \& Theuns, T. 1996, MNRAS, 282, 436

\reference{lc:96} Cole, S., \& Lacey, C. 1996, MNRAS, 281, 716


\reference{ch:64} Crampin, D.J., \& Hoyle, F. 1964, ApJ, 140, 99


\reference{dss:97} Dalcanton, J.J., Spergel, S.N, \& Summers, F.J. 1997, 
ApJ, 482, 659


\reference{d:81} Dekel, A. 1981 AA, 101, 79

\reference{ds:83} Dekel, A., \&  Schlosman  I. 
 1983, in IAU Symp. No. 100 Internal Kinematics and Dynamics of
 Galaxies, ed. E. Athanassoula (Dordrecht: Reidel). p 187

\reference{d:70} Doroshkevich, A.G. 1970, Astrofizika, 6, 581


\reference{ej:79} Efstathiou, G., \& Jones, G.J.T. 1979, MNRAS, 186, 133

\reference{eb:83} Efstathiou, G., \& Barnes, J. 1983, in Proc. 3d Moriond 
Astrophysics
Meeting, Formation and Evolution of Galaxies and Large Structures
in the Universe, ed., J.Audouze and J. Tran Thanh Van (Dordrecht:Reidel),
P. 361


\reference{els:62} Eggen, O.J., Lynden-Bell, D., \& Sandage, A.R. 1962, 
ApJ 136, 748

\reference{fe:80} Fall, S.M., \& Efstathiou, G. 1980, MNRAS, 193, 189

\reference{fpbf:93} Flores, R.,  Primack, Joel R., Blumenthal, George R., \&
Faber, S. M. 1993, ApJ 412, 443

\reference{f:70} Freeman, K.C., 1970, ApJ, 160, 811

\reference{fwde:88} Frenk, C.S., White, S.DM., Davis, M., \& 
Efsathiou, G., 1988 ApJ, 327, 507


\reference{g:00} Gardner, J. 2000,  preprint astro-ph/0006342

\reference{hp:88} Heavens, A. \& Peacock, J.A. 1988, MNRAS, 232, 339 



\reference{h:00} Hernquist, L. 1990, ApJ, 356, 359

\reference{i:66} Innanen, K.A. 1966, AJ, 71, 64

\reference{kh:97}  Klypin A. A., \&  Holtzman J. 1997, astro-ph/9712217 


\reference{art} Kravtsov, A., Klypin, A., \& Khokhlov, A.M. 1997, 
ApJS, 111, 73 

\reference{lc:93} Lacey C., \& Cole S. 1993, MNRAS, 262, 627 

\reference{lk:97} Lemson, G., \&  Kauffmann, G. 1999, MNRAS, 302, 111

\reference{lp:87} Lin, D.N.C., \& Pringle, J.E. 1987, ApJ, 320, 87L

\reference{m:63} Mestel, L. 1963, MNRAS 126, 553

\reference{mmw:98a} Mo, H.J., Mao, S., \& White, S.D.M. 1998a, MNRAS, 295, 319

\reference{mmw:98b} Mo, H.J., Mao, S., \& White, S.D.M. 1998b, MNRAS, 297L, 71

\reference{mmw:99} Mo, H.J., Mao, S., \& White, S.D.M. 1999, MNRAS, 304, 175

\reference{nfw:95} Navarro, J.F., Frenk, C., \& White, S.D.M. 1995,
 MNRAS, 275, 56
 
\reference{nfw:96} Navarro, J.F., Frenk, C., \& White, S.D.M. 1996, ApJ, 
462, 563 

\reference{nfw:97} Navarro, J.F., Frenk, C., \& White, S.D.M. 1997, ApJ, 
490, 493 (NFW)

\reference{ns:97} Navarro, J.F., \& Steinmetz, M. 1997, ApJ, 478, 13

\reference{ns:00} Navarro, J.F., \& Steinmetz, M. 2000, ApJ, 538, 477


\reference{opb:91} Olivier, S. S., Primack, J. R., \&  Blumenthal, G. R.
1991, MNRAS, 252, 102

\reference{p:69} Peebles P.J.E., 1969, ApJ, 155, 393

\reference{pdh:00} Porciani, C., Dekel, A., \& Hoffman, Y. 2000, 
in preparation 

\reference{pd:00} Porciani, C. \&  Dekel, A.  2000, 
in preparation 
 

\reference{rg} Ryden, B. S., Gunn, J. E. 1987, ApJ, 318, 15

\reference{sigad}   Sigad, Y., Kolatt, T.S., Bullock, J.S., 
Kravtsov, A.V., Klypin, A.A., Primack, J.R., 
 \& Dekel, 2000, MNRAS, submitted 


\reference{sk:99} Somerville, R.S., \& Kolatt, T.S. 1999, MNRAS, 305, 1

\reference{ssk:00} Sugerman, B, Summers, F.J., \&  Kamionkowski, M. 2000,
MNRAS, 311, 762


\reference{vdb:98} van den Bosch, F. C. 1998, ApJ, 507, 601

\reference{vdb:00}  van den Bosch, F. C. 2000, ApJ, 530, 177

\reference{vdbd:00} van den Bosch, F.C., \& Dalcanton, J.J.  2000, ApJ
 534, 146


\reference{wqsz:92} Warren, M.S., Quinn, P.J., Salmon, J.K, \& Zurek, W.H.
1992, ApJ, 399, 405

\reference{w:01} Wechsler, R. et al. 2001, in preparation

\reference{wee:98} Weil, M.L., Eke, V.R., \& Efstathiou, G. 1998, MNRAS
300, 773

\reference{wr:78} White, S.D.M., \& Rees, M.J. 1978, MNRAS, 183, 341

\reference{w:84} White, S.D.M. 1984, MNRAS, 286, 38





\reference{zn:83} Zel'dovich, Ya. B., \& Novikov, I.D. 1983, in 
Relativistic Astrophysics, ed. G. Steigman (Chicago:University of Chicago
Press), p. 384.


\reference{zqs:88} Zurek, W.H., Quinn, P.J., \& Salmon, J.K. 1988, ApJ, 330, 519

 
\end{references}
\end{document}